\documentclass{aastex63}

\newcommand\RI{FRB\,20121102A}
\newcommand\RIII{FRB\,20180916B}
\newcommand\ASKAPFRB{FRB\,20190608B}

\usepackage{amsmath}
\usepackage{bm}
\usepackage{latexsym}
\usepackage{outlines}
\usepackage{threeparttable}
\usepackage{mdframed}

\shorttitle{Intersteallar turbulence in FRB host galaxies}
\shortauthors{Simard \& Ravi}

\graphicspath{{./}{figures/}}
\begin{document}

\title{Measuring interstellar turbulence in fast radio burst host galaxies}

\email{dana.simard@astro.caltech.edu,vikram@astro.caltech.edu}

\author[0000-0002-8873-8784]{Dana Simard}
\affiliation{Cahill Center for Astronomy and Astrophysics, MC 249-17 California Institute of Technology, Pasadena, CA 91125, USA}
\affiliation{Owens Valley Radio Observatory, California Institute of Technology, 100 Leighton Lane, Big Pine, CA, 93513-0968, USA}

\author[0000-0002-7252-5485]{Vikram Ravi}
\affiliation{Cahill Center for Astronomy and Astrophysics, MC 249-17 California Institute of Technology, Pasadena, CA 91125, USA}
\affiliation{Owens Valley Radio Observatory, California Institute of Technology, 100 Leighton Lane, Big Pine, CA, 93513-0968, USA}

\begin{abstract}
Turbulence is a vital part of the interstellar medium (ISM) of galaxies, contributing significantly to galaxy energy budgets and acting as a regulator of star formation. Despite this, little is understood about ISM turbulence empirically. In the Milky Way, multiple tracers are used to reconstruct the density- and velocity-fluctuation power spectra over an enormous range of scales, but questions remain on the nature of these fluctuations at the smallest scales. Even less is known about the ISM of distant galaxies, where only a few tracers of turbulence, such as non-thermal broadening of optical recombination lines, are accessible.  We explore the use of radio-wave scattering of fast radio bursts (FRBs) to add a second probe of turbulence in extragalactic galaxies on scales many orders of magnitude smaller than those probed by emission-line widths.  We first develop the formalism to compare the scattering measures of FRBs to alternative probes of density and velocity fluctuations in the host-galaxy ISM. We then apply this formalism to three FRBs with detailed host-galaxy analyses in the literature, with the primary motivation of determining whether FRB scattering within the host galaxy probes the same turbulent cascade as the gas seen in H${\rm \alpha}$ emission. In all cases we consider, we find such an association plausible, although in one of these sources, \RI{}, the radio-scattering limit on the turbulent energy is much less constraining than the H${\rm \alpha}$ line width. We anticipate that future FRB surveys, especially those at frequencies below 1\,GHz, will find many FRBs that illuminate the small-scale properties of extragalactic ISM. 
\end{abstract}


\section{Introduction} \label{sec:intro}

Turbulence has long been thought to cause observed density fluctuations in the Galactic interstellar medium (ISM) on sub-AU to tens of parsec scales \citep{es04,se04}. Models for turbulence in the ISM, from the \citet{k41} self-similar cascade to Iroshnikov-Kraichnan \citep{i64,k65} magneto-hydrodynamic (MHD) turbulence, generally predict a power-law spectrum of density fluctuations between outer and inner length scales ($l_o$ and $l_i$ respectively). Energy is injected at the outer scale, likely by a combination of supernovae and O-type stars \citep{nf96,mk04,hjm+12,krumholz_unified_2018} and MHD instabilities that extract energy from Galactic differential rotation \citep{sb99}, and dissipated at the inner scale. Turbulence in the cold/warm neutral phases of the ISM likely causes the formation of molecular-cloud structures that are the sites of present-day star formation \citep{mo07,lla10}.

Turbulence in the warm ionized medium (WIM) has been traced for over 50 years \citep[e.g.,][]{r70} using observations of radio-wave propagation \citep{s01,hs13} and H${\rm \alpha}$ emission \citep{hdb+09}. A remarkable finding from these observations is the ``big power-law in the sky'': the concordance of several tracers of density fluctuations with a Kolmogorov spectrum ($P(k) \propto k^{-11/3}$, where $k$ is the spatial wavenumber) between outer scales $l_o\sim10^{18}$\,m and inner scales $l_i\sim10^{5}$\,m \citep{armstrong_electron_1995,chepurnov_2010}. More precisely, the outer scale has been found to vary from a few parsec in spiral arms and the inner Galaxy \citep{hgm+04} to $\sim100$\,pc in the interarm regions \citep{hbg+08}, possibly associated with the sizes of HII regions and old supernova-driven bubbles respectively. The observed inner scale likely corresponds to characteristic kinetic scales (ion inertial length or the Larmor radius) in the WIM \citep{spangler_evidence_1990}.\footnote{Recent in situ measurements of plasma-density fluctuations beyond the heliosheath by the {\em Voyager 1} spacecraft have extended the power law to scales of just 50\,m, in fact revealing an excess of power on the kinetic scales \citep{ll19}.} Power laws with similar slopes have also been observed in the densest regions of the WIM \citep{sc88,rickett_inner_2009}. Density fluctuations in the WIM on all scales, and the associated magnetic-field irregularities, are critical to our understanding of cosmic-ray transport and feedback processes in the ISM \citep{z13,hsc+21,lazarian_diffusion_2021}. Tantalizingly, it may be that the approximate  balance between the WIM, magnetic-field, and cosmic-ray pressures observed in the Galactic ISM \citep{f98} may not be universal in other galaxies \citep[e.g.,][]{b07,bbg+20}, with interesting implications for feedback in different environments. 

The above picture of WIM turbulence is complicated by instances of extreme scattering of radio sources. Radio-wave diagnostics of conditions in the WIM (and the coronal component of the ISM) can be understood in terms of the dispersion relation of a magnetized plasma \citep[e.g.,][]{rl86}: 
\begin{equation}
    \frac{c^{2}k_\lambda^{2}}{\omega^{2}} = 1-\frac{\omega_{p}^{2}}{\omega^{2}\pm\omega\omega_{B}},
    \label{eq:disp}
\end{equation}
where $c$ is the vacuum speed of light, $\omega$ is the angular frequency, $k_\lambda$ is the wavevector, $\omega_{p}=4\pi n_{e}q_{e}^{2}/m_{e}$ is the plasma frequency ($n_{e}$ is the electron number density, $q_{e}$ is the electron charge, $m_{e}$ is the electron mass), and $\omega_{B}=q_{e}B_{0}/(m_{e}c)$ is the cyclotron frequency ($B_{0}$ is the magnetic-field strength). The sign of the term in the right-hand denominator depends on the sense of circular polarization. Equation~\ref{eq:disp} implies that radio waves (\textit{a}) experience different propagation delays at different frequencies, (\textit{b}) are differentially refracted (i.e., scattered) by a spatially inhomogeneous medium, and (\textit{c}) in the case of linearly polarized radiation, trace the magnetic field strength through Faraday rotation. Extreme scattering was first observed as few-month chromatic magnification/de-magnification episodes in compact extragalactic radio sources \citep{fdj+87}. If caused by spherically symmetric plasma structures, these extreme scattering events (ESEs) are associated with small (0.1--10\,AU), dense ($n_{e}\sim10\mathrm{\,to\,}10^{3}$\,cm$^{-3}$) features that are over-pressured relative to the 8000\,K, $n_{e}\sim0.1\mathrm{\,to\,}0.5$\,cm$^{-3}$ WIM. Structures with similar properties have been identified at specific distances along the sightlines of Galactic radio pulsars \citep{smc+01,hsa+05,cks+15}, which have been found to result in highly anisotropic scattering \citep[i.e., the scattering medium has a preferred axis for ray deflections;][]{wms+04,brisken_100_2010}. Anisotropic scattering by structures at just a few parsec from the Earth is also responsible for the extreme scintillations of intra-day variable sources \citep[e.g.,][]{kjw+97}. 

Extreme radio-wave scattering is difficult to reconcile with models for turbulent cascades in the Galactic ISM \citep{se04}. In particular, the localized nature of the dominant scattering medium along the majority of pulsar sightlines \citep{bcc+04}, and the implied physical properties of the extreme scattering media, cannot be explained by standard turbulence models.  This raises serious questions about the physical nature of the observed power-law spectrum of density fluctuations, given that on $\sim$AU scales the spectrum is probed primarily by radio-wave scattering \citep{armstrong_electron_1995}. Furthermore, there are important implications for our understanding of the structure of the WIM, and of the dissipation of energy injected at the observed outer scales of the density-fluctuation spectrum. Anisotropic scattering at least is expected in models for strong turbulence caused by non-linear interactions between Alfv\'{e}n waves in the WIM \citep{gs95}. Folded corrugated sheet-like magnetic structures, motivated physically by numerical simulations of small-scale dynamos \citep{gs06}, may explain instances of extreme scattering  and solve the over-pressurization problem \citep{pk12,pl14}.

This paper is motivated by the prospect of comparing tracers of density fluctuations on multiple scales (tens of parsecs to AU) in the WIM of galaxies besides the Milky Way. By doing so, we aim to study multiple instances of density-fluctuation spectra, and thus obtain insights into their astrophysical origins.  Pulsars, one of the few types of radio sources sufficiently compact to undergo refractive scintillation, are not luminous enough for their pulsed radio emission to be detected from distant galaxies. However, fast radio bursts (FRBs) associated with specific regions within galaxies \citep[e.g.,][]{bassa_121102_2017,marcote_repeating_2020,chittidi_dissecting_2020} offer unique opportunities to probe extragalactic AU-scale density fluctuations. 

Since the discovery of FRBs over a decade ago, several hundreds of bursts from more than five hundred FRB sources have been detected.\footnote{\url{http://frbcat.org}} FRBs are dispersed by the ISM of host galaxies, the intergalactic medium (IGM), the circum-galactic medium (CGM) of intervening galaxies, and the halo and interstellar medium (ISM) of the Milky Way \citep{cc19}. Density fluctuations along FRB sightlines at any of these locations can also cause FRBs to be observed along multiple paths, resulting in an exponential distribution of burst arrival times. Rays propagating along multiple paths may coherently interfere, resulting in strong spectral modulations. Although such scattering effects are not ubiquitous in FRB samples \citep{r19}, both cases have  been observed \citep[e.g.][]{farah_frb_2018,chime_observations_2019,day_high_2020}, and in some cases attributed to the host galaxy \citep{masui_dense_2015} or the IGM (and CGM of intervening galaxies) \citep{ravi_magnetic_2016,cho_spectropolarimetric_2020}. FRB scattering promises to provide diagnostics of the characteristics of the density fluctuations in extragalactic plasmas. 

In this paper, we develop the formalism to constrain density fluctuations on multiple scales in extragalatic ISM. We do so by comparing measures of FRB scattering with optical recombination-line tracers of their host environments. We start in Section \ref{sec:turbulence} by describing the formalism we will adopt for comparing these two different tracers of turbulence.  In Section \ref{sec:location}, we discuss ways to determine the location of scattering material, in order to reasonably ensure that the FRB scattering measure is indeed probing the host galaxy environment or material immediately surrounding the FRB source, rather than the Milky Way or intervening galaxy halos.  Then, in Section \ref{sec:case_studies}, we illustrate the application of the formalism we develop to three FRBs localized to specific regions within their host galaxies with detailed optical follow-up observations. In Section \ref{sec:conclusions} we present some concluding remarks with a focus on targeting future optical and VLBI follow-up campaigns of localized FRBs for understanding extragalactic ISM properties. 

\section{Probes of turbulence on multiple scales}\label{sec:turbulence}

Studies of turbulence in the Milky Way ISM take advantage of tracers of density and velocity fluctuations on multiple scales.  Diffractive scattering of pulsars probes some of the smallest scales of density fluctuations in the Milky Way. Analyses of pulsar scattering have been used to constrain the inner scale, which is the scale at which energy is dissipated from the turbulent cascade \citep{spangler_evidence_1990, rickett_inner_2009}.  At the largest scales, gas velocities, for example from HI velocity maps, can be used to reconstruct the velocity fluctuation structure function \citep{chepurnov_extending_2010}.

Very few of these tracers are accessible at extragalactic distances.  Some galaxies of interest may be sufficiently nearby to study velocity dispersion through maps of, for example, HI emission. When we do not have sufficient spatial resolution to resolve scales below the outer scale, we can instead access the velocity dispersion of ionized gas through broadening of recombination lines.  Such a measurement is sensitive to velocity fluctuations on scales below the spatial resolution of the instrument but dominated by the largest contributing scale (i.e.,\ the smaller of the outer scale and the resolution of the instrument).  Pulsars, used to probe density fluctuations at the other end of the fluctuation spectrum in the Milky Way, are unfortunately not sufficiently bright to be routinely observed at extragalactic distances.

Diffractive scattering of FRBs probes density fluctuations at the Fresnel scale, the radius of the region in the lens plane that contributes to phase fluctuations for any ray that passes through the lensing material.  In the case studies presented in Section \ref{sec:case_studies}, the Fresnel scale is $\sim 10^9$\,m for scattering within the host galaxies.  The velocity-widths of emission lines, on the other hand, are sensitive to the outer scale of the turbulent fluctuations, $\sim 10^{18}$\,m in our case studies and orders of magnitude larger than the scales probed by scattering. 

The power spectrum of turbulent density fluctuations is typically parameterized as a power law,
\begin{equation}\label{eqn:P_n}
    P_{n}(q,l) = C_n^2(l) q^{-\beta} e^{-q^2 l_i^2}
\end{equation}
on scales smaller than the outer scale, i.e.,\ for $q<q_o$.\footnote{We have used the same definition of the inner scale, $l_i$, as \citet{macquart_temporal_2013}.  Many authors use a definition of the inner scale that is larger by a factor of $2\pi$ \citep[e.g.][]{spangler_evidence_1990,cordes_diffractive_1998}. This factor should be accounted for when comparing values from different works.}  Here $q = \frac{2\pi}{l}$ is the spatial frequency, $q_o = \frac{2\pi}{l_o}$, and $l_i$ and $l_o$ are the inner and outer scales of the density fluctuations. We will use $l$ to represent light travel distances and $d$ for angular diameter distances. The power spectrum of velocity fluctuations takes the same form
\begin{equation}
    P_{v}(q,l) = C_v^2(l) q^{-\beta} e^{-q^2 l_i^2}\;,\label{eqn:P_v}
\end{equation}
with the same inner and outer scale but a different fluctution amplitude, $C_v$.  The velocity fluctuation spectrum is proportional to the energy spectrum (per unit mass),
\begin{equation}
    E(q) = 2\pi q^2 P_v(q) \propto q^{2-\beta}\;.
\end{equation}
When $\beta > 2$, the largest scales (smallest $q$) dominate the energy in turbulence.  Turbulence in the Milky Way ISM is studied through many different tracers; the amalgamation of these tracers over 11 orders of magnitude is consistent with  Kolmogorov turbulence, for which $\beta=11/3$  \citep{armstrong_electron_1995}.  

\subsection{Density fluctuations and radio-wave scattering}

The scattering medium can be characterized by the scattering measure (SM), which is the integral of the density fluctuations along the path through the medium,
\begin{equation}\label{eqn:SM}
    \mathrm{SM} = \int C_n^2(l) \mathrm{d}l\;.
\end{equation}
\citet{macquart_temporal_2013} generalize the relation between the SM and the diffractive scale, $r_\mathrm{diff}$ (the characteristic correlation length of fluctuations in the scattering medium) for scattering at cosmological distances:
\begin{equation}\label{eqn:SM_rdiff}
    \mathrm{SM} = 
    \begin{cases} 
        \left(r_\mathrm{diff}^2 \pi r_e^2 \lambda^2 (1+z_\mathrm{lens})^{-2} l_i^{\beta-4} \frac{\beta}{4} \Gamma(-\frac{\beta}{2})\right)^{-1} & r_\mathrm{diff} < l_i \\
        \left(r_\mathrm{diff}^{\beta-2} 2^{2-\beta} \pi r_e^2 \lambda^2 (1+z_\mathrm{lens})^{-2} \beta \frac{\Gamma(-\beta/2)}{\Gamma(\beta/2)}\right)^{-1} & r_\mathrm{diff} \gg l_i\;.
    \end{cases}
\end{equation}
Here, $r_e$ is the classical electron radius and $z_\mathrm{lens}$ is the redshift of the scattering material. 

The diffractive scale is related to the observed scattering time $\tau$,
\begin{equation}\label{eqn:rdiff}
    r_\mathrm{diff} = \sqrt{ \frac{d_\mathrm{lens} d_\mathrm{lens,src}}{d_\mathrm{src}} \left(\frac{\lambda}{2\pi}\right)^2 (1+z_\mathrm{lens})^{-1} (c\tau)^{-1}}\;,
\end{equation}
where $\lambda$ is the observing wavelength.  If $r_\mathrm{diff} < l_i$,
\begin{equation}\label{eqn:SMtau}
    \tau = \frac{1}{4 \pi c} \frac{d_\mathrm{lens} d_\mathrm{lens,src}}{d_\mathrm{src}} \frac{\lambda^4 r_e^2}{(1+z_\mathrm{lens})^3} l_i^{\beta-4}\frac{\beta}{4} \Gamma\left(-\frac{\beta}{2}\right) \mathrm{SM}\;.
\end{equation}

Equations \eqref{eqn:SMtau} and \eqref{eqn:SM} allow us to use FRB scattering to constrain the amplitude of the density fluctuations at the Fresnel scale, $r_F$, given by \citet{schneider_gravitational_1992} and \citet{macquart_temporal_2013} for a cosmological lens
\begin{equation}\label{eqn:fresnel}
	r_F(z) = \sqrt{\frac{d_\mathrm{lens} d_\mathrm{lens,src} \lambda_\mathrm{obs}}{2 \pi d_\mathrm{src} (1+z_\mathrm{lens})}}\;,
\end{equation}
where $\lambda_\mathrm{obs}$ is the observing wavelength, and $d_\mathrm{lens}$, $d_\mathrm{lens,src}$ and $d_\mathrm{src}$ are the angular diameter distances between the observer and the scattering screen, the screen and the FRB source, and the observer and the FRB source respectively.  Note that $d_\mathrm{lens,src} \ne d_\mathrm{src} - d_\mathrm{lens}$; instead, in a flat Universe, 
\begin{equation}
    d_\mathrm{lens,src} = d_\mathrm{src} - \frac{1+z_\mathrm{lens}}{1+z_\mathrm{src}} d_\mathrm{lens}\;,
\end{equation}
where $z_\mathrm{src}$ is the redshift of the source \citep[][pp.\ 336-337]{peebles_principles_1993}. In this work we focus on cases where the lens is close to the source and $z_\mathrm{lens} \approx z_\mathrm{src}$, for which we write
\begin{align}
    r_F(z) &\approx \sqrt{\frac{\lambda}{2\pi} \frac{d_\mathrm{lens, src}}{1+z_\mathrm{src}}}\\
    &\approx 3.7\times10^8\,\mathrm{m}\,
    \left(\frac{f_\mathrm{obs}}{1\,\mathrm{GHz}}\right)^{-1/2}
    \left(\frac{d_\mathrm{lens, src}}{100\,\mathrm{pc}}\right)^{1/2}
    \left(\frac{1+z_\mathrm{src}}{1.1}\right)^{-1/2}\;.
\end{align}

Note from the form of equation \eqref{eqn:SMtau} that temporal scattering is greatest for screens approximately half-way between the observer and the source.  At low redshifts, scattering from the Milky Way and the FRB host galaxy is weighted approximately equally, but the additional redshift term will weight scattering in the Milky Way over scattering in the host galaxy for sources at high redshifts.

\subsection{Velocity fluctuations and dispersion}

Broadening of emission and absorption lines provides a probe of the velocity distribution in the emitting or absorbing gas. This distribution includes thermal and turbulent motions, as well as bulk motion due to, for example, gravitational infall or rotation of a disk. While bulk motion can have a more complex distribution, line broadening due to both thermal and turbulent motions is well-described by a Gaussian broadening function, and the total velocity dispersion is the quadrature sum of the thermal and turbulent velocity dispersions in the absence of bulk motion.  In this work, we will use the H$\alpha$ recombination line as our velocity tracer. We assume that the H${\rm \alpha}$-emitting regions do not have significant bulk motions, and we account for only the turbulent and thermal velocity distributions.

Information on the velocity fluctuation spectrum is present in the turbulent contribution to the velocity dispersion.  The velocity power spectrum captures the mean-square fluctuations at a scale $q$, and so the variance of the velocity is given by the integral of the power spectrum over all (but the 0$^\mathrm{th}$) scales:\footnote{Normally this includes the zeroeth mode: $\langle x^2 \rangle = \int_o^\infty P_x (q) dq$.  We are not including the zeroeth mode as we only consider power on scales below the outer scale.  As a result, we measure $\int_{q_0>0}^{\infty} P_x(q) dq = \langle x^2 \rangle - \langle x \rangle^2 = \sigma_x$.}
\begin{align}
    \sigma^2_{v,nt} &= \int_{q_o}^\infty d^3\mathbf{q} P_{ v}(\mathbf{q})\label{eqn:sigmav} \\
    &= 2\pi C_v^2 q_i^{3-\beta}\, \Gamma\left(\frac{3-\beta}{2}, \left(q_o l_i\right)^2\right) \;.\label{eqn:sigma_v_Cv_lo}
\end{align}
Since $P_{v}(q) \propto q^{-\beta}$, we see that the integral in equation \eqref{eqn:sigmav} is dominated by large spatial scales when $\beta > 2$. (In Kolmogorov turbulence, $\beta =11/3 \approx 3.7$). When $\beta < 2$, the smallest scales dominate the velocity dispersion and therefore the energy.

Using the relation $\Gamma(s+1, x) = s \Gamma(s, x) + x^s e^{-x}$, we can write
\begin{align}
    \Gamma\left(\frac{3-\beta}{2}, \left(q_o l_i\right)^2\right) &= \frac{2}{\beta-3}\left(q_o l_i\right)^{\frac{3-\beta}{2}} e^{-\left(q_o l_i \right)^2}-\frac{2}{\beta-3} \Gamma\left(\frac{5-\beta}{2}, \left(q_o l_i\right)^2\right)\;.
\end{align}
For $3 < \beta < 5$ and $q_o l_i \ll 1$, this simplifies to
\begin{equation}
    \Gamma \left(\frac{3-\beta}{2}, \left(q_o l_i\right)^2\right) \approx \frac{2}{\beta-3} \left(q_o l_i\right)^{3-\beta}\;.
\end{equation}
and we find
\begin{equation}
    \sigma_v^2 \approx \frac{2(2\pi)^{4-\beta}}{\beta-3}C_v^2 l_o^{\beta-3}\;,
\end{equation}
which more immediately shows the dependence of the velocity dispersion (and the energy) on the outer scale.  

\subsection{Relating the density and velocity power spectra}

If we can relate the amplitude of the velocity fluctuation power spectrum to that of the density power spectrum, we can calculate the turbulent velocity dispersion in the ionized medium from the FRB SM.   This assumes:  (\textit{a}) the plasma probed by FRB scattering is part of the same turbulent cascade as the gas emitting in H${\rm \alpha}$; and (\textit{b}) the turbulent fluctuations follow a power-law spectrum between the scales probed by FRB scattering (the Fresnel scale of the lens, $\sim 10^8$\,m) and the velocity dispersion (the outer scale).  If we equate the outer scale to the size of knots of H$\alpha$ emission close to FRB sources in their host galaxies, we infer outer scales of $\sim 100\,\mathrm{pc} \simeq 3\times10^{18}$\,m, similar to the outer scale in the Milky Way \citep{armstrong_electron_1995}; see Section \ref{sec:case_studies} for further details. If the velocity dispersion anticipated from scattering disagrees with the observed velocity dispersion, at least one of these assumptions must be incorrect. 

If the amplitude of the turbulent fluctuations inferred from FRB scattering is much larger than the amplitude inferred from emission line widths, this may suggest that FRBs originate from anomalously highly-scattered environments in their host galaxies. This would be analogous to the case of the Crab pulsar; the Crab nebula is one of the most highly-scattered environments that we observe in the Milky Way. The Crab pulsar exhibits variable scattering timescales up to 600\,$\mu$s at 610\,MHz \citep{mckee_temporal_2018}.  The variability of this scattering time indicates that scattering is occurring close to the source, likely within the nebula itself. Adopting a distance of 0.5\,pc between the source and scattering screen\footnote{\citet{main_mapping_2018} argue that optically-emitting filaments in the Crab Nebula at this distance are the source of the the temporal radio-wave scattering.}, and assuming a Kolmogorov spectrum with an inner scale of 1000\,km, this corresponds to a SM of $\sim$300\,kpc\,m$^{-20/3}$ (equation \ref{eqn:SMtau}).  Similarly, underscattering of FRBs compared to expectations from H${\rm \alpha}$ line widths implies that the observed FRBs are in fact not associated with the nearby H${\rm \alpha}$-emitting regions.  Associations (or non-associations) of FRBs with star-forming regions are especially impactful for understanding the progenitors of FRB sources.

An alternative explanation for a discrepancy in the turbulent energies inferred from radio-wave scattering and optical line widths is that the assumption that the turbulent power spectrum follows a simple power law with $\beta=11/3$ is incorrect.  Inferred departures from $\beta=11/3$ are not unprecedented in the Milky Way. The evolution of some  pulsar scattering timescales over frequency ($\tau \propto \nu^{-\alpha}$) suggests a departure from Kolmogorov turbulence; pulsar scattering exponents \citep[of $\alpha \sim 3\,\mathrm{to}\,4$, compared to the Kolmogorov value of $\alpha=4.4$; e.g.,][]{geyer_scattering_2017} suggest $\beta > 11/3$, and indicate increased power on large scales compared to Kolomogorov turbulence, although it is likely that this effect arises from anisotropic scattering at discrete screens.\footnote{\citet{geyer_scattering_2017} find that fitting anisotropic models can increase the inferred $\alpha$ for some pulsars, while \citet{oswald_thousand_2021} find a mean value of $\alpha=4.0\pm0.6$ for a sample of distant pulsars, which are expected to be scattered by many screens with different orientations leading to an overall isotropic scattering kernel.} VLBI observations towards the Galactic center indicate increased power in turbulence on small scales compared to a Kolmogorov spectrum. From the size of the scattered disk of Sgr A* (sensitive to the smallest fluctuations, at the inner scale, $\sim800\,$km) and refractive substructure in the disk (sensitive to scales of $10^{13}-10^{14}$\,cm) \citet{johnson_scattering_2018} infer a slope of the density fluctuation power spectrum of $\beta = 3.38$. Heightened power near the inner scale is also observed in the density spectrum of the solar wind \citep{coles_solar_1991}, in this case due to a flattening of the power spectrum at small scales.  \citet{johnson_scattering_2018} note that a similar flattening is consistent with their Sgr A* measurements.   Obtaining additional tracers at intermediate scales would allow us to measure the  index of the power spectrum and discern breaks or changes in the power-law form. We examine the feasibility of probing density and velocity fluctuations at additional scales for FRB host galaxies in Section \ref{sec:conclusions}.

Simulations of isothermal turbulence \citep[e.g.][]{passot_density_1998, konstandin_new_2012} show a linear relationship between the standard deviation of the density distribution, $\sigma_\rho$, and the rms Mach number, $\mathcal{M} = v_\mathrm{rms}/c_s$, of the gas,
\begin{equation}
    \sigma_\rho = b \mathcal{M} \langle \rho \rangle \;,
\end{equation}
where $v_\mathrm{rms}$ is the rms velocity, $c_s$ is the speed of sound in the gas, $\langle \rho \rangle$ is the volume-weighted mean density, and $b$ is a constant.  \citet{konstandin_new_2012} find that, in the supersonic regime, $b=1$ and $b=1/3$ for compressive and solenoidal forcing respectively.  A mixture of both types of forcing is likely in the ISM \citep{federrath_comparing_2010}; for simplicity we choose a fiducial value $b=1$.

In analogy to equation \eqref{eqn:sigma_v_Cv_lo}, we can write
\begin{equation}
    \sigma_{\delta n_e}^2 = (2\pi)^{4-\beta} C_n^2 q_o^{3-\beta}\frac{2}{\beta-3}\;.
\end{equation}
This allows us to relate the amplitudes of the velocity and density fluctuations
\begin{equation}\label{eqn:C_v_C_n}
    C_{v}^2 = \left(\frac{c_s}{b \langle n_e \rangle} \right)^2 C_n^2 \;,
\end{equation}
where $\langle n_e \rangle$ is the average electron number density, and also provides a relationship between the velocity dispersion, $\sigma_v$, and the amplitude of the velocity perturbation power spectrum.

The emission measure, $\mathrm{EM}$, calculated from the H${\rm \alpha}$ emission line, probes the square of the electron density along the line-of-sight, $\mathrm{EM} = \int n_e^{2} \mathrm{d}l$.   The EM in a clumpy medium, with electron density fluctuations within the clumps, is related to the mean electron density by \citep[][equation B3]{cordes_radio_2016}:
\begin{equation}
    \mathrm{EM} = \frac{\zeta (1+\epsilon^2)}{f_f} \langle n_e \rangle^2 L_{\mathrm{EM}}\;.
\end{equation}
We adopt fiducial values of $\epsilon=1$ and $\zeta=2$ corresponding to complete modulation of the electron density inside the clouds and 100\% variation of the average $n_e$ between clouds for consistency with \citet{tendulkar_host_2017}. $L_\mathrm{EM}$ is the path length through the medium that contributes to the EM and $f_f$ is the filling factor of the clouds within the medium.  

Using this in equation \eqref{eqn:C_v_C_n}, we can write,
\begin{align}
    C_v^2 &= \left(\frac{c_s}{b}\right)^2 \frac{\zeta (1+\epsilon^2)}{f_f f_r} \frac{L_\mathrm{EM}}{\mathrm{EM}} C_n^2\\
    &= \left(\frac{c_s}{b}\right)^2 \frac{\zeta (1+\epsilon^2)}{f_f f_r} \frac{\mathrm{SM}}{\mathrm{EM}}\;,
\end{align}
where we have used $\mathrm{SM} = C_n^2 L_\mathrm{SM}$ and assumed that the path length that contributes to the SM $L_\mathrm{SM} = f_r L_\mathrm{EM}$, where $f_r \le 1$.  This accounts for the possibility that the FRB source is embedded in the scattering medium, in which case a shorter path length will contribute to the SM than to the EM.

We can then write the velocity dispersion as 
\begin{align}\label{eqn:v_from_SM}
    \sigma_v^2 &= \frac{2 (2\pi)^{4-\beta}}{\beta-3} l_o^{\beta-3} \left(\frac{c_s}{b}\right)^2 \frac{\zeta (1+\epsilon^2)}{f_f f_r} \frac{\mathrm{SM}}{\mathrm{EM}}\\
        &= \frac{2 (2\pi)^{4-\beta}}{\beta-3} l_o^{\beta-3} \left(\frac{c_s}{b\,n_e}\right)^2 \frac{4\pi c}{f_r} \frac{d_\mathrm{src}}{d_\mathrm{lens} d_\mathrm{lens,src}} \frac{(1+z_\mathrm{lens})^3}{\lambda^4 r_e^2} l_i^{4-\beta}\frac{4}{\beta \Gamma(-\beta/2)} \tau \;.
\end{align}
For Kolmogorov ($\beta=11/3$) turbulence,
\begin{align}
    \sigma_v^2 &= 3 (2\pi)^{1/3} l_o^{2/3} \left(\frac{c_s}{b}\right)^2 \frac{\zeta (1+\epsilon^2)}{f_f f_r} \frac{\mathrm{SM}}{\mathrm{EM}}\label{eqn:sigmav_SM_1}\\
    \begin{split}
    &= \left(1.4\times10^5\,\mathrm{km}\,\mathrm{s}^{-1}\right)^2
    \left(\frac{l_o}{150\,\mathrm{pc}}\right)^{2/3}
    \left(\frac{c_s}{10\,\mathrm{km}\,\mathrm{s}^{-1}}\right)^{2}
    \left(\frac{b}{1}\right)^{-2}
    \left(\frac{\zeta}{2}\right)
    \left(\frac{f_f}{0.1}\right)^{-1}
    \left(\frac{f_r}{0.5}\right)^{-1}
    \left(\frac{1+\epsilon^2}{2}\right)\times\\
    &\,\,\,\,\,\left(\frac{\mathrm{SM}}{10^5\,\mathrm{kpc}\,\mathrm{m}^{-20/3}}\right)
    \left(\frac{\mathrm{EM}}{600\,\mathrm{pc}\,\mathrm{cm}^{-6}}\right)^{-1}\;.\label{eqn:sigmav_SM_2}
    \end{split}
\end{align}

With a measurement of the SM (from radio observations of the FRB) and EM (from optical observations of the host galaxy), combined with an estimate of the path length through the star-forming region, we can directly compare the velocity dispersion predicted from the SM and EM to the velocity dispersion from optical recombination-line widths. We will use a fidicual sound speed of $c_s = \sqrt{\frac{kT}{\mu}}=10$\,km\,s$^{-1}$, for hydrogen gas at a temperature of $10^4$\,K. Increased gas metallicities increase the mean molecular weight $\mu$ and decrease the sound speed, but the sound speed is most sensitive to the temperature of the medium. An independent measure of the temperature would allow us to constrain the sound speed more accurately. 

If instead we assume that the optical emission lines and the FRB SM are probing the same turbulent spectrum, we can assert that the velocity dispersions inferred from these two tracers must be equal and constrain the nature of the turbulent cascade.  Particularly, as $\sigma_v^2  \propto l_o^{\beta-3} l_i^{4-\beta} n_e^{-2}$, we can constrain the inner and outer scales and the index of the turbulent power spectrum through this method.  We can also vary the electron density, $n_e$; the inference of the average electron density from the EM of the emitting gas depends on the clumpiness of the electrons in the gas, parameterized through the filling factor of ionized gas in the medium and the electron density variations between and within the ionized clouds.  Additional constraints on $\langle n_e \rangle$ translate to constraints on these clumpiness parameters.

\subsection{Energetics of turbulence at the scattering screen}\label{sec:turbulence_radio}

In addition to testing whether FRB scattering and the velocity dispersion of the ionized gas are probing the same plasma and the same turbulent cascade, we can use both observations to estimate the turbulent energy in the region.  In this section, we develop the formalism for relating the SM and the velocity dispersion to the turbulent energy.  As above, we will consider isotropic turbulence with density and velocity fluctuations described by equations \eqref{eqn:P_n} and \eqref{eqn:P_v}.  Since we are considering scattering originating in the host galaxy, the path length through the scattering plasma is only a small percentage of the entire path length between the observer and source, and we will use a thin-screen formalism to relate the FRB SM and the amplitude of the density fluctuation power spectrum. 

The energy per unit mass can be calculated from the integral of the the kinetic energy in the velocity fluctuations,
\begin{align}
    E &= \frac{1}{2} \int \mathrm{d}^3{\bf q} P_v(q)\\
    &= 2\pi \int_{q_o}^{\infty} C_v^2 q^{2-\beta} e^{-\frac{q^2}{q_i^2}}\\
    &= \frac{(2\pi)^{4-\beta}}{\beta-3} l_o^{\beta-3} \left(\frac{c_s}{b}\right)^2 \frac{\zeta (1+\epsilon^2)}{f_f f_r} \frac{\mathrm{SM}}{\mathrm{EM}}\\
    &=\frac{(2\pi)^{4-\beta}}{\beta-3} l_o^{\beta-3} \left(\frac{c_s}{b\,n_e}\right)^2 \frac{4\pi c}{f_r} \frac{d_\mathrm{src}}{d_\mathrm{lens} d_\mathrm{lens,src}} \frac{(1+z_\mathrm{lens})^3}{\lambda^4 r_e^2} l_i^{4-\beta}\frac{4}{\beta \Gamma(-\beta/2)} \tau \;.
\end{align}
For Kolmogorov turbulence,
\begin{align}
    E&= \frac{3}{2} (2\pi)^{1/3} l_o^{2/3} \left(\frac{c_s}{b}\right)^2 \frac{\zeta (1+\epsilon^2)}{f_f f_r} \frac{\mathrm{SM}}{\mathrm{EM}}\;\label{eqn:turb_energy}\\
    \begin{split}
    &= 2.0\times10^{53}\,\mathrm{erg}M_\odot^{-1}
    \left(\frac{l_o}{150\,\mathrm{pc}}\right)^{2/3}
    \left(\frac{c_s}{10\,\mathrm{km}\,\mathrm{s}^{-1}}\right)^2
    \left(\frac{b}{1}\right)^{-2}
    \left(\frac{\zeta}{2}\right)
    \left(\frac{f_f}{0.1}\right)^{-1}
    \left(\frac{f_r}{0.5}\right)^{-1}
    \left(\frac{1+\epsilon^2}{2}\right)\times\\
    &\,\,\,\,\,\left(\frac{\mathrm{SM}}{10^5\,\mathrm{kpc}\,\mathrm{m}^{-20/3}}\right)
    \left(\frac{\mathrm{EM}}{600\,\mathrm{pc}\,\mathrm{cm}^{-6}}\right)^{-1}\;.
    \end{split}
\end{align}

Equation \eqref{eqn:turb_energy} can also be written in terms of the non-thermal velocity dispersion  as
\begin{equation}\label{eqn:turb_energy_sigmav}
    E = \frac{1}{2} \sigma_{v, nt}^2 \;.
\end{equation}
Using this equation, we can also estimate the energy in the turbulence from the non-thermal gas velocity dispersion measured from ionized emission lines. 

\section{Determining the distance to the scattering plasma}\label{sec:location}

As the radio emission from FRBs propagates towards us, it interacts with the intervening ionized plasma.  This includes any plasma associated with the FRB source itself, such as a pulsar wind nebula, supernova remnant or other ejecta from a compact binary merger \citep[e.g.][]{piro_dispersion_2018, straal_dispersion_2020, kundu_impact_2020, zhao_dispersion_2021}, the host galaxy ISM, including star-forming and HII regions, the IGM and the CGM of intervening galaxies, halos of the Milky Way and intervening galaxies \citep{ocker_constraining_2021}, and the ISM of the Milky Way.  

The effects of the Milky Way ISM are constrained observationally by pulsar scattering and dispersion.  From these measurements, models of the electron density distribution within the Milky Way ISM, such as the NE2001 model \citep{cordes_ne2001_2002,cordes_ne2001_2003} and the YMW16 model \citep{yao_new_2017}, have been developed.  These include features, such as spiral arms, within the Milky Way and predict the effects of dispersion and scattering in the Milky Way ISM on both galactic and extragalactic radio sources.\footnote{We use the term scattering to refer to various effects due to multi-path propagation, including scintillation, temporal broadening, and angular broadening.}  Using these predictions, we can determine if the scattering observed towards any particular FRB predominantly originates from the Milky Way.

A similar tomography is yet to be realized for the IGM, although FRB dispersion and scattering measurements hold promise for mapping the electron content of the IGM out to high redshifts \citep[e.g.][]{mcquinn_locating_2014, zheng_probing_2014, ravi_measuring_2019, caleb_constraining_2019}. \citet{macquart_census_2020} present the first DM-$z$ relation constructed from five FRBs with localizations and host-galaxy associations, out to redshifts of $z\sim0.5$; the trend of this relation describes the mean electron content of the IGM. FRBs that fall outside of this relation may intercept a deficit of ionized electrons due to voids along the light-travel path, or an abundance due to heightened dispersion in the host galaxy, and filaments or CGM of intervening galaxies along the light-travel path.  Detailed analyses of individual FRBs have also yielded rich information about scattering in the IGM \citep{ravi_magnetic_2016,cho_spectropolarimetric_2020,simha_disentangling_2020}, the halos and CGM of intervening galaxies \citep{prochaska_low_2019,connor_skewer_2020} and the host galaxy \citep{masui_dense_2015}.

One of the challenges of using FRBs to study density fluctuations in intervening material is attributing extragalactic scattering to the appropriate material along the line-of-sight.  This is especially key for in-depth studies of individual systems, where we do not have the opportunity to marginalize over many lines-of-sight and host galaxies.  One way to constrain the distance to the scattering material is to use the relationship between angular and temporal broadening, which depends on the distances between the observer and the source and scattering screen.  Scattering angles can and have been measured by modelling the scintillation pattern on the Earth (to localize interstellar scattering towards both pulsars \citep{rickett_interstellar_2014, reardon_modelling_2019, reardon_precision_2020, marthi_scintillation_2020} and AGN \citep[e.g.][]{bignall_observations_2007, oosterloo_extreme_2020, wang_askap_2021}), directly through VLBI \citep[to localize interstellar scattering towards pulsars;][]{brisken_100_2010, smirnova_radioastron_2014, popov_distribution_2016, shishov_interstellar_2017, fadeev_revealing_2018},  and using size constraints from the presence of scintillation in two-screen scattering systems \citep[applied to FRB scattering to place an upper limit on the source-screen distance;][]{masui_dense_2015, farah_frb_2018}.  

For scattering of emission from a source at an angular diameter distance $d_\mathrm{src}$ from the observer by a single thin screen at an angular diameter distance $d_\mathrm{lens}$ and redshift $z_\mathrm{lens}$ from the observer, the temporal broadening $\tau$, angular broadening $\theta$ and scintillation bandwidth $\Delta \nu$ are related by (see Appendix \ref{sec:theta})
\begin{equation}\label{eqn:tau}
    \tau \simeq \frac{1}{c} \frac{d_\mathrm{src} d_\mathrm{lens}}{d_\mathrm{lens,src}}\theta^2 (1+z_\mathrm{lens})
\end{equation}
\citep{schneider_gravitational_1992} and 
\begin{equation}
    2 \pi \Delta \nu \tau = C_1\;\label{eqn:c1}
\end{equation}
\citep{cordes_diffractive_1998}, where $C_1$ is a constant that depends on the distribution of the scattering material and the density fluctuations within the scattering material.  For isotropic Kolmogorov turbulence in a thin screen, $C_1 = 0.957$ \citep{cordes_diffractive_1998}.  For consistency, we will work with the scattering delay, $\tau$, when considering extragalactic scattering, but equation \eqref{eqn:c1} can be used to convert from the scintillation bandwidth $\Delta \nu$ to the delay $\tau$.

Even if the redshift of the source is known from association of the FRB with a host galaxy, there remains a degeneracy between the distance to the scattering material, the angular broadening of the source and the scattering timescale.  To determine the location of the scattering screen using the observed temporal scattering or scintillation, we need to first constrain the broadening angle.

\subsection{Constraining the broadening angle through VLBI}\label{sec:vlbi}

One might first approach this problem hoping to directly measure the broadening angle through VLBI. \citet{brisken_100_2010} have shown that this can be used to measure the brightness distribution of scattered pulsars at 100\,$\mu$as resolution, but angular broadening due to scattering outside of the Milky Way is a much smaller effect.

The rms scattering angle is related to the diffractive scale by (see Appendix \ref{sec:theta}):
\begin{align}\label{eqn:theta_rdiff}
    \theta &\simeq \frac{1}{r_\mathrm{diff}}\frac{d_\mathrm{lens,src}}{d_\mathrm{src}} \frac{\lambda}{2\pi} \frac{1}{1+z_\mathrm{lens}}\\
    &\simeq 70\,\mathrm{nas}\,
    \left(\frac{r_\mathrm{diff}}{10^7\,\mathrm{cm}}\right)^{-1}
    \left(\frac{d_\mathrm{lens,src}}{100\,\mathrm{pc}}\right)
    \left(\frac{d_\mathrm{src}}{400\,\mathrm{Mpc}}\right)^{-1}
    \left(\frac{f_\mathrm{obs}}{300\,\mathrm{MHz}}\right)^{-1}
    \left(\frac{1+z_\mathrm{lens}}{1.1}\right)^{-1}\;,
\end{align}
and to the delay by
\begin{align}\label{eqn:theta_tau}
    \theta &\simeq \sqrt{c \tau \frac{d_\mathrm{lens,src}}{d_\mathrm{lens} d_\mathrm{src}}\frac{1}{1+z_\mathrm{lens}}}\\
    &\simeq 70\,\mathrm{nas} 
    \left( \frac{\tau}{25\,\mathrm{ms}}\right)^{1/2}
    \left(\frac{d_\mathrm{lens,src}}{100\,\mathrm{pc}}\right)^{1/2}
    \left(\frac{d_\mathrm{src}}{400\,\mathrm{Mpc}}\right)^{-1} 
    \left(\frac{1+z_\mathrm{lens}}{1.1}\right)^{-1/2}
\end{align}
Here, we use a fiducial observing frequency of $300$\,MHz, as the best constraints are placed at low frequencies, where the decrease in the angular resolution of VLBI ($\propto f_\mathrm{obs}$) is offset by the increased angular broadening of the source ($\propto f_\mathrm{obs}^{-2.2}$). Resolving angles on this scale would require baselines of $\gtrsim 2.8\times10^9$\,km at 300\,MHz, $4\times10^5$ times the radius of the Earth.

Even if the capabilities existed to resolve this broadening, a further complication arises in detecting this broadening in the presence of additional scattering in the Milky Way and IGM.  The $d_\mathrm{lens,src}/(1+z_\mathrm{lens})$ term in equation \eqref{eqn:theta_rdiff} more strongly weights scattering occurring closer to the observer.  As a result, angular broadening in the Milky Way will likely obscure any from scattering within the host galaxy. Scattering in the IGM is also likely to obscure angular broadening contributed by scattering in the host galaxy.  The CGM of an intervening galaxy consisting of a fog-like two-phase medium, as proposed by \citet{mccourt_characteristic_2018}, is expected to impart an angular broadening  $\sim15\,\mu$as for a source at redshift $z \sim 1$ \citep{vedantham_radio_2019}, two orders of magnitude larger than the expected broadening from scattering in the host galaxy. \citet{macquart_temporal_2013} estimate angular broadening from a clumpy IGM, in which the density fluctuations of clouds throughout the IGM follow a log-normal probability distribution, to be $\lesssim 100$\,$\mu$as at $z\lesssim1$.\footnote{Both \citet[][equation (13)]{macquart_temporal_2013} and \citet[][equation (17)]{vedantham_radio_2019} use an incorrect scaling of the scatter-broadening angle with redshift: $\theta_\mathrm{scat}=\frac{d_\mathrm{lens,src}}{d_\mathrm{src}}\frac{\lambda}{2\pi r_\mathrm{diff}}$. Compared to equation \eqref{eqn:theta_rdiff}, this is incorrect by a factor of $(1+z_\mathrm{lens})$. We have ignored this factor, which is order unity for redshifts $z_\mathrm{lens} \lesssim 1$, in our comparison between angular broadening due to scattering in the IGM and in the host galaxy.  See Appendix \ref{sec:appendixB} for details.} 

\subsection{Constraining the broadening angle by modelling the scintillation pattern}

The spatial coherence scale of the scintillation pattern, $l_d$, depends directly on the rms scattering angle, $\theta$:
\begin{equation}
    l_d = \frac{\lambda}{2\pi\theta}
\end{equation}
\citep[][equation (6)]{cordes_diffractive_1998}\;.  With single-epoch single-dish measurements, only the decorrelation timescale, which depends on both the speed of the scintillation pattern and spatial coherence scale, can be measured. Multi-station observations or observations over the orbital period of the Earth or, for Galactic sources in binary systems, the source, can break this degeneracy and even constrain any anisotropy in the scattering (including both the axial ratio and position angle of the scattered disk). See \citet{bignall_rapid_2006} and \citet{reardon_modelling_2019} for details and  \citet{fadeev_revealing_2018}, \citet{simard_comparison_2019}, and \citet{marthi_scintillation_2020} for further examples.

Measuring the spatial scintillation patterns of FRBs is not feasible in most cases, due to the one-off nature of non-repeating sources, and the long waiting times often observed between repeat bursts, although an upper limit on the scintle size can be placed if different scintillation patterns are observed at different stations for the same burst.  It is also plausible that some systems will be discovered with high repetition rates and duty-cycles, from which we can constrain the scintillation pattern. Superluminal motion of FRB emission regions would similarly act to constrain the scintillation pattern size \citep{simard_galactic}. 

Using the rms scattering angle from equation \eqref{eqn:theta_tau}, we find
\begin{align}
    l_d \approx \frac{\lambda}{2\pi}\left(\frac{c\tau}{1+z_\mathrm{lens}} \frac{d_\mathrm{lens,src}}{d_\mathrm{lens} d_\mathrm{src}}\right)^{-1/2}\;.
\end{align}
For lensing occurring close to the source,
\begin{align}
    l_d  \approx 4.6\times10^{8}\,\mathrm{km}\,
    \left(\frac{f_\mathrm{obs}}{1405\,\mathrm{MHz}}\right)^{-1}
    \left(\frac{\tau}{1\,\mathrm{ms}}\right)^{-1/2}
    \left(\frac{d_\mathrm{lens,src}}{100\,\mathrm{pc}}\right)^{-1/2}
    \left(\frac{d_\mathrm{src}}{400\,\mathrm{Mpc}}\right)
    \left(\frac{1+z_\mathrm{lens}}{1.1}\right)^{1/2}\;.
\end{align}
Again, baselines between stations orders of magnitude larger than the radius of the Earth are needed to constrain the angular size of the scattered disk using this technique.

\subsection{Constraining the broadening angle through scattering in the Milky Way}\label{sec:galactic_scattering}

\begin{figure}
\centering
 \includegraphics[width=0.9\textwidth]{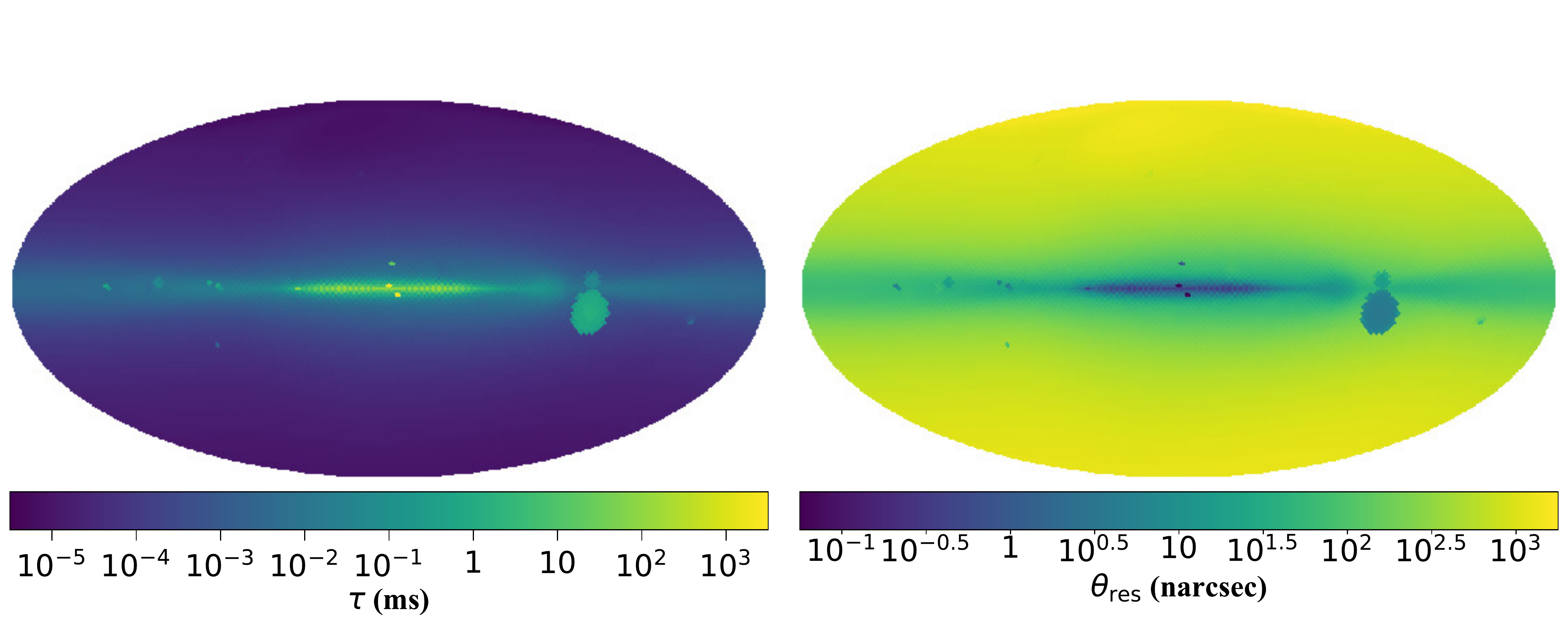}\\
 \includegraphics[width=0.9\textwidth]{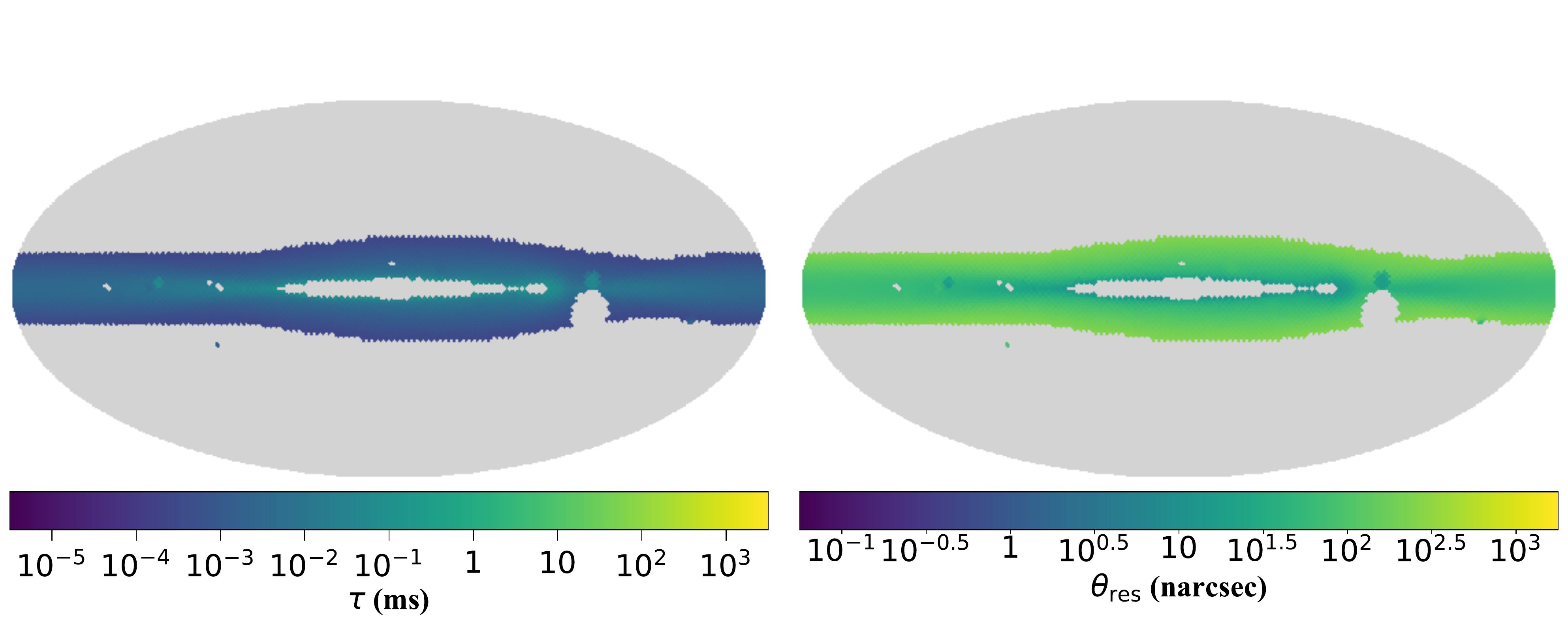}
\caption{{\bf Top:} The expected scattering timescale due to scattering in the Milky Way at 1.405\,GHz from the NE2001 model of the electron distribution in the Milky Way (left) and the resolution of the Milky Way scattering screen assuming a distance to the screen of 1\,kpc (right).  {\bf Bottom:} The same plot with regions masked for which the DSA-110 (with 250-MHz bandwidth) would be insensitive to galactic scintillation or $\gtrsim1$\,ms of extragalactic scattering.  Note that the DSA-110 is best suited for detecting scintillation from Milky Way scattering screens with resolutions of 0.1 to 1 $\mu$as, and is not sensitive to scintillation along Milky Way sightlines with high scattering (such as towards the Galactic center) or with very low scattering (such as those at high Galactic latitudes).  \label{fig:scattering}}
\end{figure}

Alternatively, we can take advantage of scattering in the Milky Way ISM to constrain the angular broadening at an extragalactic screen.   In order for scintillation due to scattering in the Milky Way to be observed, the source must be unresolved by the Milky Way scattering screen.\footnote{While the scattering within the Milky Way may be distributed through the ISM or occur at multiple screens, we will approximate Milky Way scattering as occurring at a thin screen, as the distance through the Milky Way is only a small fraction of the total distance to the FRB source.}  The angular resolution of the Milky Way scattering screen, $\theta_\mathrm{res,MW}$, is given by 
\begin{equation}
    \theta_\mathrm{res,MW} = \frac{\lambda}{\theta_\mathrm{MW}d_\mathrm{MW}}\;,
\end{equation}
where $\lambda$ is the observing wavelength and $\theta_\mathrm{MW}$ and $d_\mathrm{MW}$ are respectively the broadening angle due to scattering at and the distance to the Galactic scattering screen.  In Fig.\ \ref{fig:scattering}, we show the scintillation bandwidth and resolution of the Milky Way scattering screen calculated from the NE2001 model \citep{cordes_ne2001_2002}, which predicts the scattering timescale or scintillation bandwidth induced by scattering within the Milky Way, at a frequency of 1.405\,GHz (assuming $\tau \propto f_\mathrm{obs}^{-4.4}$ and a distance to the Milky Way screen of 1\,kpc).  Note that the scintillation bandwidth varies by many orders of magnitude between sightlines towards the Galactic Center and sightlines at high Galactic latitudes; however, instruments will be sensitive to only a small range of scintillation timescales determined by the frequency resolution and bandwidth.  In the lower panel of Fig.\ \ref{fig:scattering},  we have demonstrated this effect by excluding regions where it would be difficult to detect Galactic scintillation or 1-ms extragalactic scattering with the DSA-110 instrument, an 110-dish array dedicated to FRB detection and localization at the Owens Valley Radio Observatory.  This includes high Galactic latitudes where the Galactic scintillation bandwidth is $>10$\,\% of the observing bandwidth of the DSA-110, and low Galactic latitudes where the Galactic scattering timescale is $>10$\,\% of our fiducial 1-ms extragalactic scattering delay and would impair our ability to measure the scattering delay imparted by extragalactic plasma.\footnote{We do not consider the effect of the channel width on detecting scintillation with very small scintillation bandwidths; with the retention of baseband data, the DSA-110 and other upcoming surveys will be able to search for Galactic scintillation after detection by resampling the baseband data with finer frequency resolution.}

When scintillation is observed that is consistent with our expectations for scattering in the Milky Way from the NE2001 model, the presence of scintillation indicates that the FRB is unresolved by the Galactic screen and we can constrain the angular broadening of any extragalactic scattering to be 
\begin{equation}
    \theta_\mathrm{EG} < \theta_\mathrm{res,MW}\;.
\end{equation}
When scattering is observed, this angular limit can be combined with the scatter-broadening timescale, $\tau_\mathrm{EG}$, to place a limit on the separation between the scattering screen and the source:
\begin{align}
    d_\mathrm{src} - d_\mathrm{EG} &< \frac{\theta_\mathrm{res,MW}^2 d_\mathrm{src} d_\mathrm{EG}}{c \tau_\mathrm{EG}} (1+z_\mathrm{EG})\;,
\end{align}
where $z_\mathrm{EG}$ and $d_\mathrm{EG}$ are the redshift of and angular diameter distance to the extragalactic scattering screen.  When the scattering is occurring close to the source, we can write:
\begin{align}
    d_\mathrm{src} - d_\mathrm{EG} &< \frac{\theta_\mathrm{res,MW}^2 d_\mathrm{src}^2}{c \tau_\mathrm{EG}} (1+z)\\
    &< 32\,\mathrm{kpc}
    \left(\frac{\theta_\mathrm{res,MW}}{100\,\mathrm{nano-arcsec}}\right)^2
    \left(\frac{\tau_\mathrm{EG}}{1\,\mathrm{ms}}\right)^{-1}
    \left(\frac{d_\mathrm{src}}{1\,\mathrm{Gpc}}\right)^2 
    \left(\frac{1+z_\mathrm{EG}}{1.325}\right)\;.
\end{align}
This argument has been used by \citet{masui_dense_2015} to associate scattering and a high rotation measure to the host galaxy of FRB\,110523 and by \citet{farah_frb_2018} to place the material scattering FRB\,170827 by 4.1\,$\mu$s to within 60\,Mpc of the source.\footnote{This latter source-screen distance limit is underestimated by a factor of $(2\pi)^2$ compared to our method; this arises due to differences in the treatment of the resolution limit of the Milky Way screen.} We will adopt frequency-dependencies of the scattering timescales and broadening angles of $\tau_\mathrm{EG} \propto f_\mathrm{obs}^{-4.4}$ and $\theta_\mathrm{res,MW} \propto f_\mathrm{obs}^{-1.2}$, so the critical lens-source distance that we are sensitive to, $d_\mathrm{crit}$, is $\propto f_\mathrm{obs}^{4.6}$  and there is a strong preference for observations at low frequencies.  However, at very low frequencies scattering from within the Milky Way may mask any extragalactic scattering, limiting our ability to constrain the extragalactic scattering timescale.

For an unchanging scattering timescale with distance (i.e.,\ ignoring any redshift-evolution of the turbulent properties of FRB host galaxies as well as the redshift dependence in equation \eqref{eqn:tau}), the critical distance between the extragalactic lens and source for which the Milky Way scattering screen resolves the further scattering screen is  $\propto d_\mathrm{src}^2 (1+z_\mathrm{EG})$, assuming the extragalactic lens is close to the source. This is maximized (assuming cosmological parameters from \citet{planck_cosmological_2015}, including $H_0 = (67.8\pm0.9)$\,km\,s$^{-1}$\,Mpc$^{-1}$ and $\Omega_M = 0.308\pm0.012$) at $z=3.8$, at which a Galactic screen resolution of $<44$\,nano-arcseconds is needed to resolve a distance $d_\mathrm{lens,src} = 50\,$kpc for an extragalactic scattering timescales of 1\,ms.  At redshifts below 0.2 (for which we can obtain detailed observations of the host galaxy to identify star-forming regions and the location of the FRB within the host), a Galactic screen resolution of $<190$\,nano-arcseconds is required to resolve a screen-source separation of 50\,kpc when $\tau=1\,$ms.  To understand the prospects of multi-tracer studies of turbulence in FRB host galaxies with current and upcoming FRB surveys, we  estimate the fraction of FRBs detected by these surveys at redshifts $<0.2$ and seen through parts of the Milky Way with sufficient angular resolution to resolve a 50\,kpc lens-source distance.  We  focus on the DSA-110 FRB survey, which will enter a commissioning and early science phase in summer of 2021.  In Appendix \ref{sec:appendixB}, we present similar estimates for the current ASKAP CRAFTS survey as well as CHORD and DSA-2000, future instruments in their design phases. In all cases, we use frbpoppy\footnote{\url{http://davidgardenier.com/frbpoppy/}} \citep{gardenier_frbpoppycode_2019} to estimate the redshift distribution of detected FRBs, as described in Appendix \ref{sec:appendixB}.

We use the NE2001 model to obtain the resolution of the Galactic screen, and adopt a minimum declination of $-30$\,deg for the DSA-110. We find that 1.2\% and 17.4\% of the sky visible to the DSA-110 has sufficient resolution to be sensitive to a screen-source separation of 50\,kpc at all redshifts and at $z\leq0.2$, respectively, for a FRB with 1-ms scattering at 1.405\,GHz.  If the FRB population traces stellar mass density in the Universe, then approximately 1 in 4 of FRBs detected by the DSA-110 will be at $z<0.2$. If the DSA-110 spends equal time searching all areas of the sky, then approximately 1 in 24 of DSA-110 FRBs will meet these conditions (low redshift, viewed through a sufficiently-resolving part of the Milky Way) to undertake an in-depth study of turbulence and scattering in the host galaxy as described here. The final number of DSA-110 sources for which this study can be done also depends on the distribution of scattering in host galaxies, which we do not attempt to model here.

\begin{figure}
\centering
\includegraphics[width=0.75\textwidth]{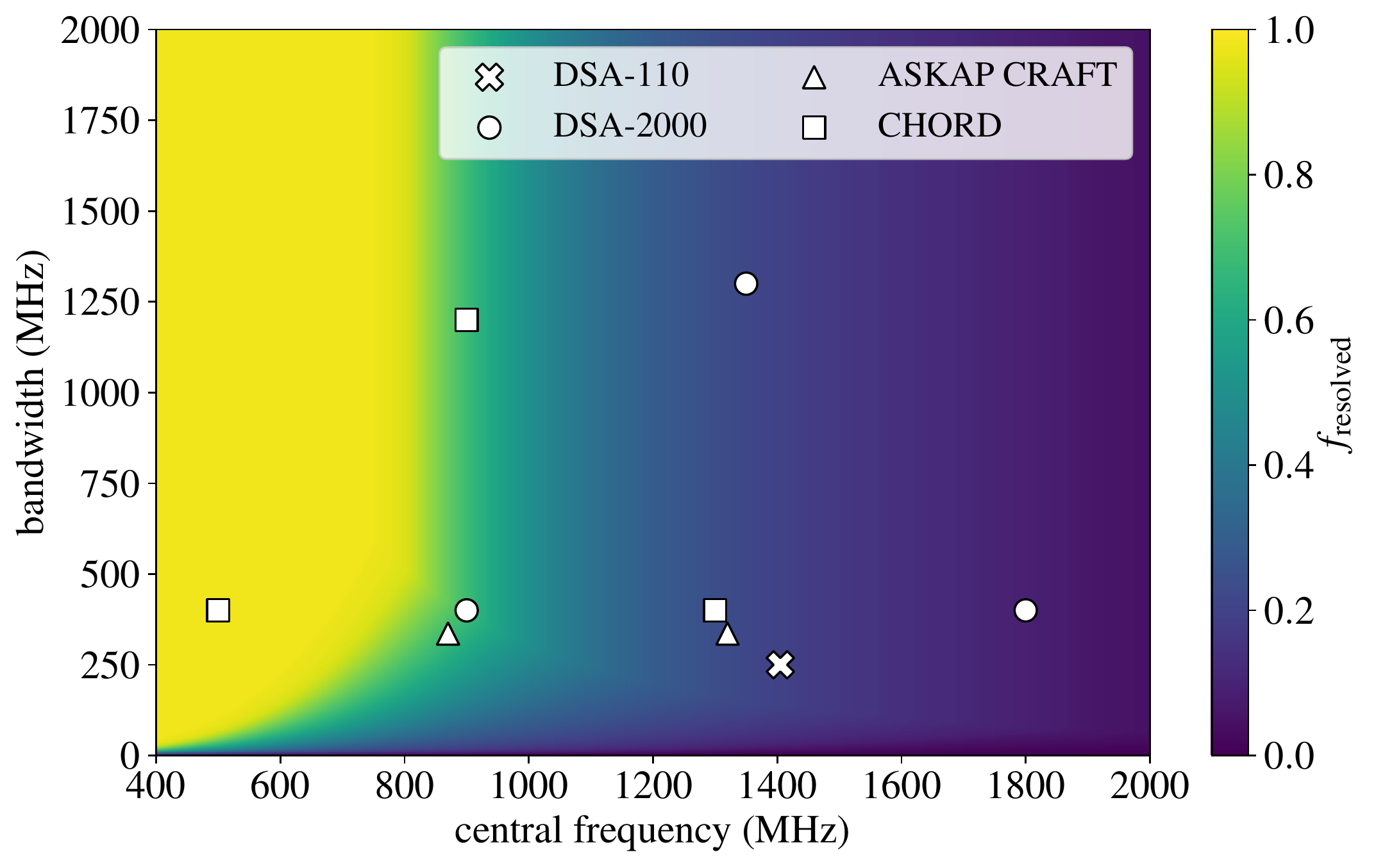}
\caption{The fraction of sightlines through the galaxy for which a 50-kpc source-screen distance could be resolved, as a function of central observing frequency and observing bandwidth.  We have adopted a fiducial scattering timescale for scattering in the host galaxy of 1\,ms at 1.405\,GHz, and assume that the timescales for both extragalactic scattering and scattering in the Milky Way (calculated using the NE2001 model at 1\,GHz) are presumed to scale as $f_\mathrm{obs}^{-4.4}$. The DSA-110 survey, with a central frequency of 1.405\,GHz and a bandwidth of 250\,MHz is indicated by the white x.  In Appendix \protect\ref{sec:appendixB}, we also consider the coherent-search mode of the ASKAP CRAFT survey (triangles), the DSA-2000 (circles) and CHORD (squares).  For each of these, we consider different central frequencies and bandwidths within the instrument's capabilities. \label{fig:surveys}}
\end{figure}

In Fig.\ \ref{fig:surveys}, we show how the fraction of the sky for which the Milky Way has sufficient resolution changes with central frequency and bandwidth of the instrument.  We clearly see two effects: an increased bandwidth and lower central frequency (Galactic scintillation bandwidths are narrower at lower frequencies) allow detection of Galactic scintillation over a larger part of the sky, increasing the likelihood of observing an FRB for which we can localize the extragalactic scattering material using a two-screen model; and at lower frequencies the required resolution of the galactic screen decreases due to the increased fiducial scattering  timescale (and broadening angle) at the extragalactic screen. The required resolution scales as $f_\mathrm{obs}^{-2.2}$ (assuming $\tau_\mathrm{EG} \propto f_\mathrm{obs}^{-4.4}$), while the resolution of the Galactic scattering screen scales as $f_\mathrm{obs}^{-1.2}$ (again, we assume $\tau_\mathrm{MW} \propto f_\mathrm{obs}^{-4.4}$), so that at lower frequencies the required resolution is met over more of the sky. 

In all cases, we consider FRBs that are scattered to $>1$\,ms at 1.405\,GHz (the center frequency of the DSA-110 survey), and scale this scattering with $\tau \propto f_\mathrm{obs}^{-4.4}$.  At the lowest frequencies, this results in scattering timescales $\sim100$\,ms.  Recent results from the CHIME/FRB collaboration suggest that there may be a substantial undetected population of highly-scattered FRBs, but that their current detection pipeline is biased against events with scattering timescales above 10\,ms \citep{chime_catalog_2021}.  Searching for FRBs with large scattering timescales is computationally challenging with current standard single-pulse search algorithms, but further development of search algorithms, including image-based searches, may enable discoveries of FRBs with high scattering timescales.

\begin{figure}
\centering
\includegraphics[width=0.75\textwidth]{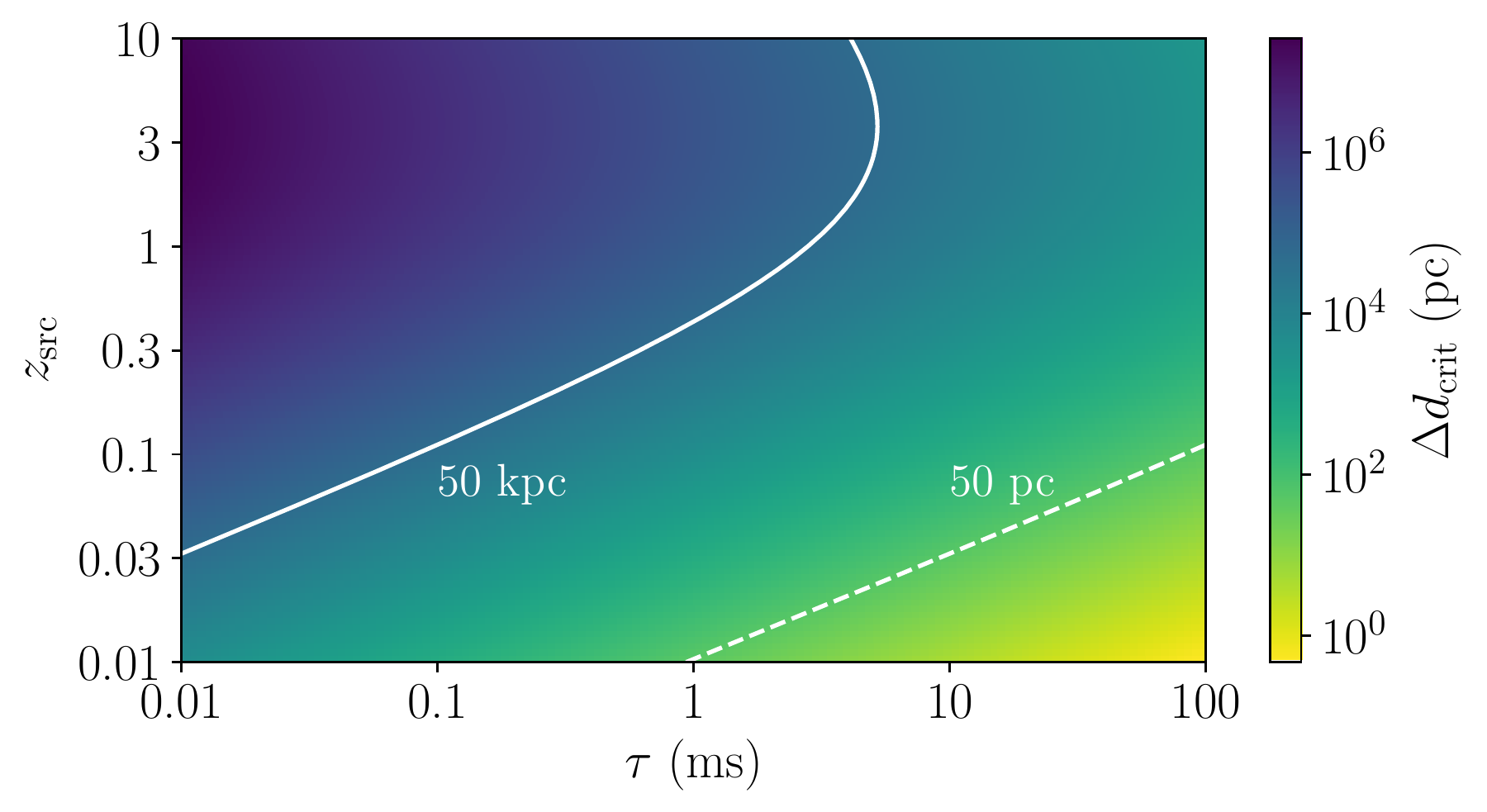}
\caption{The critical source-screen separation, $\Delta d_\mathrm{crit}$, below which scintillation due to Galactic scattering with a resolution of 100\,nano-arcseconds will resolve the scattering disk on the extragalactic screen, derived for sources at a cosmological redshift $z_\mathrm{src}$ with an extragalactic scattering delay of $\tau$. The white solid contour labels the critical distance of 50\,kpc, for which extragalactic scattering could reasonably be associated with the host galaxy, while the white dashed contour labels a critical distance of 50\,pc.
\label{fig:dcrit}}
\end{figure}

One will also encounter cases where no scintillation is observed from the Milky Way.  Depending on the uncertainty in NE2001 expectations along the sightline (which might be somewhat assuaged by proximity to nearby pulsars), one may be able to place a lower limit on the angular scatter-broadening in these cases. The lack of scintillation where expected indicates that the burst is sufficiently angular-broadened by extragalactic scattering to be resolved by the Galactic screen. When combined with an observed temporal broadening timescale, this can allow one to rule out scattering within the host galaxy of the FRB.  However, one must diligently confirm the properties of Galactic scattering towards the FRB in these cases, as some sightline-to-sightline or temporal variations in scattering (for example, extreme scattering events) are not captured in the NE2001 model.  

Additional potential scattering locations include the IGM and the CGM of intervening galaxies.  Studies of FRB scattering at these locations hold promise for furthering our understanding of the baryon content and distribution in the IGM and CGM \citep{macquart_temporal_2013}.  Recently, scattering theory has been applied to FRB observations to constrain the distribution of cold gas in the halos of intervening galaxies \citep{prochaska_low_2019,cho_spectropolarimetric_2020} and turbulence in the ionized IGM \citep{ravi_magnetic_2016}.  Statistical studies of FRB DMs will likely also prove fruitful for revealing the structure of the IGM \citep[e.g.][]{macquart_frb_2018, macquart_census_2020, xu_probing_2020}, including, for example, constraining the era of helium reionization \citep[e.g.][]{caleb_constraining_2019}. In this paper, however, we focus specifically on scattering in the ISM of FRB host galaxies.

\section{Case studies}\label{sec:case_studies}

\begin{deluxetable}{cccc}
\tablecaption{Constraints from radio and optical observations used to infer the turbulent properties of the host galaxies and host star-forming regions (SRs) of repeating FRBs. The SMs, and some EMs, are derived in this work. \label{tbl:observations}}
\tablehead{\colhead{}  & \colhead{\ASKAPFRB{}} & \colhead{\RIII{}} & \colhead{\RI{}} } 
\startdata
\multicolumn{4}{c}{\it Radio constraints}\\
$\tau$ (ms)            & 3.3$\pm$0.2$^{a}$ & $<$1.6$^{d}$           & $<$9.6$^{f}$ \\
$f_\mathrm{obs}$ (MHz) & 1200$^{a}$         & 350$^{d}$              & 500 \\
$z$                    & 0.1178$^{b}$      & 0.0337$\pm$0.002$^{e}$ & 0.19273$\pm$0.00008$^{g}$ \\
$\mathrm{SM}$ (kpc\,m$^{-20/3}$) & 240$\pm$130 & $<$0.7 & $<(1.0\pm0.5)\times10^5$ \\
\hline
\multicolumn{4}{c}{\it Optical constraints on associated star-forming regions}\\
H${\rm \alpha}$ flux ($10^{-16}$ erg\,cm$^{-2}$\,s$^{-1}$) & --          & 1.002$\pm$0.008$^{e}$ & 2.6 \\
Size of SR (pc)       & 640$\pm$320$^{c}$ & 190$^{h}$ & 1320 \\
$\sigma_v$ & 15$\pm$0.55 km\,s$^{-1}$\,$^{c}$ & $290\pm50$\,km\,s$^{-1}$ & 91.8$\pm$1.6\,km\,s$^{-1}$ \\
EM (pc\,cm$^{-6}$) & 222$\pm$12$^{c}$ & 310$\pm$140 & 510$\pm$160 \\
\hline
\multicolumn{4}{c}{\it Assumptions}\\
$d_\mathrm{lens,src}$ (pc) & 90$\pm$50 & 100$\pm$50 & 660$\pm$330\\
$l_o$ (pc)                 & 150 & 150 & 1320 \\
\enddata
\tablenotetext{a}{\citet{day_high_2020}}
\tablenotetext{b}{\citet{macquart_census_2020}}
\tablenotetext{c}{ \citet{chittidi_dissecting_2020}; EM is inferred by \citet{chittidi_dissecting_2020} from the measured H${\rm \beta}$ flux assuming a H${\rm \alpha}$ to H${\rm \beta}$ flux ratio of 3.95.}
\tablenotetext{d}{\citet{chime_r3lowfreq_2020}}
\tablenotetext{e}{\citet{marcote_repeating_2017}}
\tablenotetext{f}{\citet{chime_r1detection_2019}}
\tablenotetext{g}{\citet{tendulkar_host_2017}}
\tablenotetext{h}{\citet{tendulkar_60_2020}; where errors are not given we have assumed 20\% uncertainty}
\tablenotetext{i}{\citet{bassa_121102_2017}}
\end{deluxetable}

\begin{deluxetable}{ccc|cc}
\tablecaption{Turbulent energy inferred from measurements of FRB scattering and optical line widths, along with estimates of the scale probed by each method.  \label{tbl:results}}
\tablehead{
    \colhead{}  & 
    \multicolumn{2}{c}{From scattering} &
    \multicolumn{2}{c}{From line widths} \\
    \colhead{}  & \colhead{Scale probed (m)} & \colhead{Energy (erg\,$M_\odot^{-1}$)} & \colhead{Scale probed (m)} &
    \colhead{Energy (erg\,$M_\odot^{-1}$)}
} 
\startdata
\ASKAPFRB{} & $3.3\times10^8$ & $3.3\times10^{49}$\,to\,$2.0\times10^{51}$ & 4.6$\times10^{18}$ & $2.2\times10^{45}$ \\
\RIII{} & 5.5$\times10^8$ & $<4\times10^{48}$ & 4.5$\times10^{18}$ & $8\pm3\times10^{47}$ \\
\RI{} &  1.3$\times10^9$ & $<3\times10^{49}$ & $4\times10^{19}$ & 8.4$\pm0.3\times10^{46}$ 
\enddata
\end{deluxetable}

We will focus on three FRBs that have been localized to H$\alpha$-emitting regions within their host galaxies: \RI{} \citep{bassa_121102_2017}, \RIII{} \citep{marcote_repeating_2020}, and \ASKAPFRB{} \citep{chittidi_dissecting_2020}.  Of these three sources, only \ASKAPFRB{} exhibits temporal scattering inconsistent with expectations from the Milky Way \citep{day_high_2020}; both \RI{} \citep{chime_r1detection_2019} and \RIII{} \citep{chime_r3lowfreq_2020} have stringent upper limits on their extragalactic scattering timescales.  For all three sources, optical imaging and spectroscopic observations have been used to characterize the host galaxies and FRB-source local environments.  Here, we consider these three sources in turn, and compare their observed scattering (or scattering limits) to constraints on the local environment from optical observations. Table \ref{tbl:observations} lists relevant constraints from both optical host-galaxy follow-up and radio FRB observations, as well as assumptions on the scattering geometry and the scales in the turbulent media.  

We will assume, for the H$\alpha$-emitting gas in all three host galaxies, a temperature of $10^4$\,K with a sound speed of 10\,km\,s$^{-1}$, and turbulence that follows a Kolmogorov spectrum with $\beta=11/3$ and an inner scale of 1000\,km, consistent with \citet{cordes_radio_2016}.  We note that the inner scale is poorly constrained even in the Milky Way.  The multi-tracer analysis of \citet{armstrong_electron_1995}, which probes down to scales of $\sim10^7$\,m, is consistent with an inner scale $\lesssim 10^8$\,m.  Detailed pulsar scattering \citep{rickett_inner_2009} studies and interferometric studies of anglar broadening and image wander \citep{spangler_evidence_1990} suggest the inner scale may be $\sim 100$\,km. However, as discussed in Section \ref{sec:intro}, evidence that localized, anisotropic scattering may dominate pulsar scintillation \citep[e.g.][]{putney_multiple_2006,brisken_100_2010} raises doubt that the plasma at tiny scales probed by pulsars is indeed part of the same isotropic turbulent cascade seen at larger scales. To obtain physical scales at the host galaxies from angular ones, we use cosmological parameters from \citep{planck_cosmological_2015}.

\begin{figure}
\centering
\includegraphics[width=0.5\textwidth]{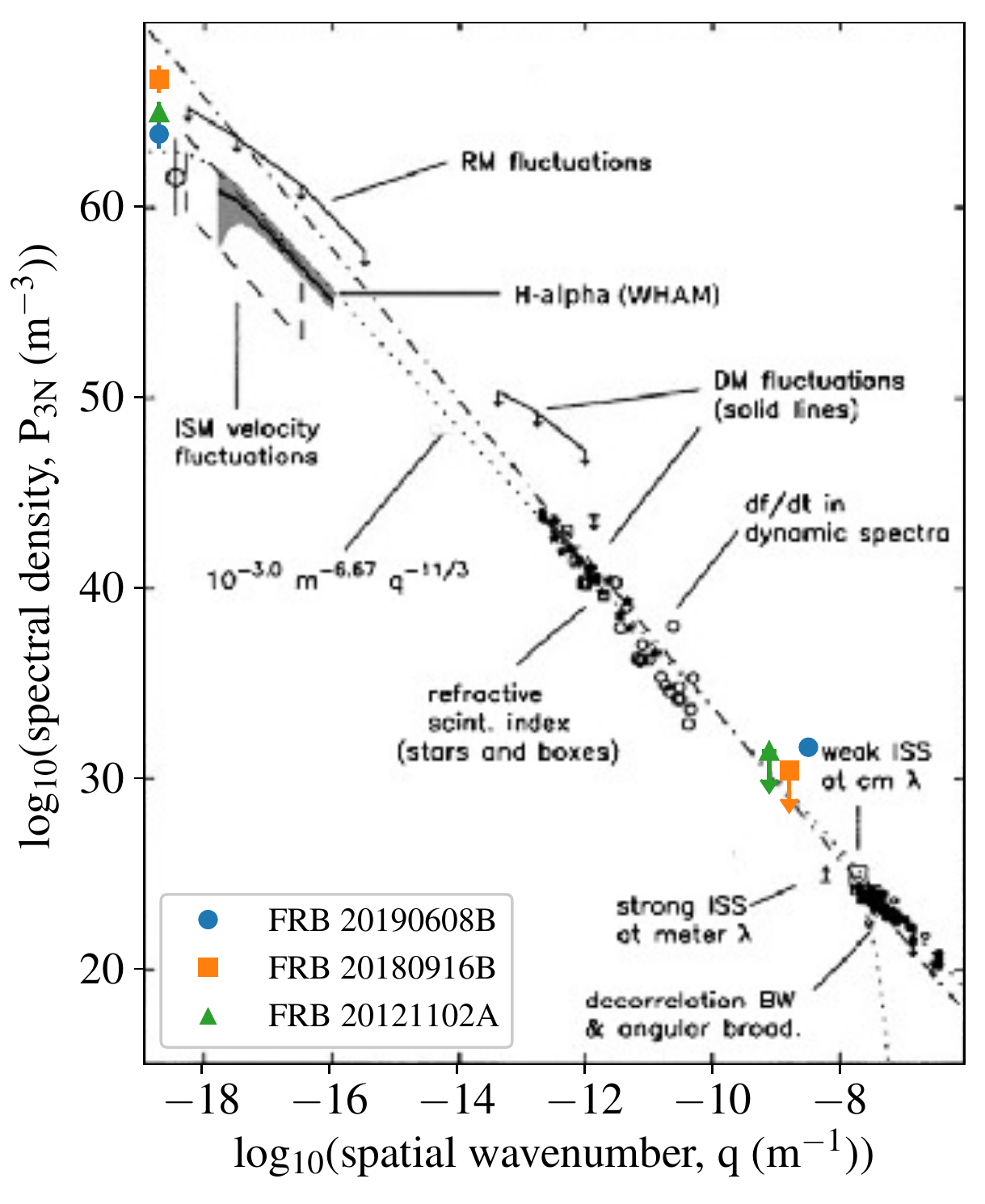}
\caption{The ``power law in the sky" measured in the Milky Way reproduced from \citet{chepurnov_extending_2010} and including the original data of \citet{armstrong_electron_1995} (black and white) with points from this study of extragalactic galaxies with FRB scattering and H$\alpha$ line widths over plotted in color.  Blue circles, orange squares, and green triangles show results for the host galaxies of \ASKAPFRB{}, \RIII{} and \RI{} respectively.  Points at large $q$ (right hand of the plot) are from scattering measurements, while points at low $q$ (left hand of the plot) are from non-thermal velocity dispersions measured from H$\alpha$ line widths. Note that for consistency with \citet{armstrong_electron_1995}, we adopt a definition of the spatial wavenumber $q=1/l$ in this figure. Where the points are not upper limits, errorbars are shown that represent uncertainties determined from the assumed uncertainty on the lens-source screen distance and the measured uncertainties on the SM, non-thermal velocity dispersion and electron density.  These error bars do not include uncertainty in the electron density, sound speed of the gas, outer scale, or the spectral index of the turbulent cascade, all of which should ultimately be considered.
\label{fig:powerlaw}}
\end{figure}

In Fig.\ \ref{fig:powerlaw}, we compare the power on the outer and Fresnel scales measured from the burst scattering timescales and emission line widths for these three FRBs and their local environments to the ``power law in the sky" in the Milky Way.  We note two major differences between studies in the Milky Way and our study: in the Milky Way, tracers are present at intermediate scales that aren't probed in our study of FRB host galaxies; and in the Milky Way tracers are typically averaged over multiple sightlines and different tracers can be acquired for different sightlines.  In this study, we are considering two tracers that probe approximately the same region of the FRB host galaxy.

\subsection{\ASKAPFRB{}}\label{sec:ASKAP}

\ASKAPFRB{} was discovered and localized by the Australian Square Kilometer Array Pathfinder (ASKAP) \citep{macquart_census_2020} and associated with a host galaxy at redshift $z=0.1178$.  The host galaxy is a spiral galaxy with stellar mass $M_\star \approx 10^{10.4}M_\odot$ and star formation rate $\mathrm{SFR} \approx 1.2 M_\odot\,\mathrm{yr}^{-1}$ \citep{bhandari_host_2020}. High time-resolution analysis of the radio burst yields a scattering timescale of $\tau = 3.3 \pm 0.2$\,ms at a frequency of 1.28\,GHz \citep{day_high_2020}, significantly longer than expected from scattering in the Milky Way (34\,ns according to the NE2001 model).\footnote{We assume $\tau \propto f_\mathrm{obs}^{-4.4}$, consistent with \citet{cordes_ne2001_2002}.} \citet{day_high_2020} also measure the spectral index of the scattering timescale, and find $\tau \propto f_\mathrm{obs}^{-\alpha}$, where $\alpha = 3.5\pm0.9$.

Here we assume that the scattering observed in \ASKAPFRB{} originates in its host galaxy.  A detailed analysis of the foreground environment by \citet{simha_disentangling_2020} finds no foreground galaxy within 2.5\,Mpc in the SDSS catalogue, as well as no foreground galaxy in the 1\,arcmin Keck Cosmic Web Imager (KCWI) field used to study the local environment of \ASKAPFRB{}.  They simulate the cosmic web using SDSS galaxies within 400 arcminutes of the line-of-sight to the host galaxy of \ASKAPFRB{} and find it implausible that the foreground is responsible for the temporal scattering of the FRB. \citet{day_high_2020} observe structure in the spectrum of \ASKAPFRB{} that may be Galactic scintillation, but for which they are unable to measure a decorrelation bandwidth.  A screen in the Milky Way with the NE2001 scintillation bandwidth of 5.4\,GHz at 1200\,MHz has a resolution of only 2\,$\mu$as, and a confirmation of scintillation in the Milky Way would allow a localization of the extragalactic scattering material to within 320\,kpc of the source. While we cannot claim that the scattering definitively originates in the host galaxy, we will continue with this case study to demonstrate the power of our analysis framework.

\citet{chittidi_dissecting_2020} examine the local environment of \ASKAPFRB{} in detail using HST WFC3/UVIS imaging and Keck/KCWI integral field spectroscopy.  The HST images revealed that the positional localization of \ASKAPFRB{} coincides with a knot of star formation in one of the spiral arms of the host galaxy.  While they do not observe the H${\rm \alpha}$ line with KCWI, they infer the H${\rm \alpha}$ flux of the star-forming knot from the H${\rm \beta}$ flux, assuming the flux ratio follows expectations for HII regions, and calculate $\mathrm{EM} = 222 \pm 12$\,pc\,cm$^{-6}$. In the KCWI H${\rm \beta}$ maps, \citet{chittidi_dissecting_2020} measure a velocity dispersion of $\sigma_v = 15 \pm 0.55$\,km\,s$^{-1}$ within a region including both the burst localization and the star-forming knot.

In the following, we derive properties of the ISM local to the source of \ASKAPFRB{} from the optical data in hand.  We subtract a thermal width of $\sigma_t = 9.1\pm2.7$\,km\,s$^{-1}$ \citep{osterbrock_2006} (corresponding to a temperature of $(1.0\pm0.5)\times 10^4$K) from the velocity dispersion in quadrature to find a non-thermal velocity dispersion of $\sigma_{nt} = 11.9\pm2.8$\,km\,s$^{-1}$. We ignore the natural broadening of the line, which is three orders of magnitude smaller than the thermal broadening. We estimate the size of the star-forming region to be $\approx 1\pm0.5$ spaxel in the KCWI images, or $640\pm320$\,pc at an angular diameter distance of 453\,Mpc.  As the host galaxy is face-on (\citet{chittidi_dissecting_2020} measure an inclination angle of $37\pm3\,\deg$), if this region were spherical it would extend above the thin disk of the galaxy, which \citet{chittidi_dissecting_2020} take to be $\sim 150$\,pc.  We estimate the line-of-sight extent to be $\sim 150\,\mathrm{pc}\,\cos^{-1}(37\,\deg)$, and take $L=190\pm90\,\mathrm{pc}$. We adopt the scale height as the outer scale, $l_o=150\,\mathrm{pc}$.  From the EM and the line-of-sight extent of the region, we infer an electron density between 0.08 and 0.85\,cm$^{-3}$, assuming a filling fraction between 0.1 and 1, $\zeta=2$ and $\epsilon=1$. This would contribute a DM to \ASKAPFRB{} in the observer frame\footnote{$\mathrm{DM}_\mathrm{obs} = \langle n_e \rangle L_\mathrm{SM} (1+z)^{-1}$, where $L_\mathrm{SM} = L_\mathrm{EM} f_r$.} of $\sim10-140$\,pc\,cm$^{-3}$, consistent with the total host contribution of $137\pm43$\,pc\,cm$^{-3}$ \citep{chittidi_dissecting_2020}.

The radio observations of \ASKAPFRB{} can be used to predict the non-thermal velocity dispersion in the ionized gas, assuming a Kolmogorov turbulent fluctuation spectrum, using equation \eqref{eqn:v_from_SM}.  In addition to measurements of the SM, EM, and outer scale, equation \eqref{eqn:v_from_SM} requires estimates of the sound speed, clumpiness of the ionized gas (given by $f_f$, $\epsilon$ and $\xi$), and the path length of the FRB through the gas. We will adopt our fiducial values of $\zeta=2$ and $\epsilon=1$.  As shown in equations \eqref{eqn:rdiff} and \eqref{eqn:SMtau}, the observed scattering time $\tau$ is related to the SM through the diffractive scale, $r_\mathrm{diff}$, the characteristic size of fluctuations on the scattering screen. For \ASKAPFRB{}, and assuming a screen-source distance $d_\mathrm{lens,src} = 90\pm50$, $r_\mathrm{diff} = 6.0\pm1.6\times10^6$\,cm and  $\mathrm{SM}=240\pm120\,\mathrm{kpc}\,\mathrm{m}^{-20/3}$.  The scattering probes fluctuations at the Fresnel scale, equation \eqref{eqn:fresnel}, $\sim 3 \times 10^8$\,m.

For $f_r=1$ and $f_f=0.1$, we calculate from the radio observations an expected velocity dispersion of $(8.2\pm2.1)\times 10^4$\,km\,s$^{-1}$, many orders of magnitude larger than the observed value of $11.9\pm2.8$\,km\,s$^{-1}$.  Varying $f_r$ between 0.5 and 1, and $f_f$ between 0.1 and 1, we find a range for the velocity dispersion of $1.9-14.6\times10^4$\,km\,s$^{-1}$.  The energies calculated from the radio scattering observations and the observed velocity dispersion from H${\rm \beta}$ KCWI maps (calculated from equations \eqref{eqn:turb_energy} and \eqref{eqn:turb_energy_sigmav}) are given in Table \ref{tbl:results}.  The energy from the optical line width is a few orders of magnitude smaller than the energy inferred from the scattering timescale, assuming that both of these are probing the same turbulent fluctuation spectrum within the H$\alpha$ region.

As the energies are separated by only a couple of orders of magnitude, we can attempt to reconcile them by varying the inner and outer scales. An outer scale six orders of magnitude smaller than our adopted value (i.e., $l_o \sim 30$\,AU) is needed to reconcile the two energies, holding the inner scale at 1000\,km.  It is difficult to motivate energy injection on such a small scale, and therefore such a small outer scale seems unlikely.  From equations \eqref{eqn:turb_energy} and \eqref{eqn:SM_rdiff} we see that $E\propto l_i^{1/3}$; however, this applies only when the inner scale is larger than the diffractive scale  (see \eqref{eqn:SM_rdiff}), which, from the scattering timescale of \ASKAPFRB{}, we calculate to be $6.0\pm1.6\,\times\,10^{6}$\,cm.  Varying the inner scale can only decrease the energy by a factor of $\sim 10$ due to this flattening in the SM function against the inner scale.  Alternatively, the electron density ($E \propto n_e^{-2}$, see equation \eqref{eqn:turb_energy}) would need to be 70 times larger, for which the region would contribute a DM between 440 and 9960\,pc\,cm$^{-3}$, much larger than the estimated host DM and even exceeding the total DM of $338.7\pm0.5$\,pc\,cm$^{-3}$ \citep{macquart_census_2020}.

A final point of uncertainty is the scaling of the power spectrum over the orders of magnitude that separate the scales probed by these two tracers.  We have assumed a Kolmogorov scaling, $\beta = 11/3 \approx 3.7$, but with a slightly shallower power law, $\beta \sim 3.2$ the energies probed by FRB scattering (at $3.3\times10^8$\,m) and the H${\rm \beta}$ velocity dispersion (at $4.6\times10^{18}$\,m) are consistent. While this is a small change in the power-law index, the energy difference at small scales is $10^5$ larger than what one would extrapolate from the velocity dispersion measurement assuming $\beta=11/3$.  Alternatively, this large excess power on small scales could be due to a break or bump in the power spectrum. Indeed, this is put forth by \citet{johnson_scattering_2018} as an explanation for the increased power on small scales in the fluctuation spectrum towards the Milky Way Galactic Center.  For isotropic turbulence, the scattering index $\alpha$ measured by \citet{day_high_2020} corresponds to a power-law index $\beta = 2\alpha (\alpha-2)^{-1}$ of $\beta=4.7\pm1.6$, consistent with the $\beta \sim 3.2$ power-law index inferred from comparing fluctuations on large and small scales and suggesting that the scattering index may be constant across the intermediate scales.  
To distinguish between these cases, additional tracers of the fluctuation spectrum or the power law index are necessary.  

\subsection{\RIII}\label{sec:R3}

\RIII{} is a repeating source discovered by the CHIME/FRB Collaboration  \citep{chime_r3disc_2019} and localized to an emission region in a massive spiral galaxy at redshift $z=0.0337\pm0.002$ \citep{marcote_repeating_2020}. \RIII{} is also the first FRB to be detected below $400$\,MHz \citep{chime_r3lowfreq_2020,pilia_r3sardinia_2020}; this low-frequency detection provides a stringent upper limit on the extragalactic scattering timescale of $<1.6$\,ms at $350$\,MHz \citep{chime_r3lowfreq_2020}.  Recently, a detection of \RIII{} with LOFAR at 110 to 188\,MHz \citep{pastor_chromatic_2020, pleunis_lofar_2020} shows a scattering tail with $\tau \sim 50$\,ms at 150\,MHz, consistent with the upper limit set by \citep{chime_r3lowfreq_2020} and with expectations for galactic scattering from the NE2001 model.  Any scattering in the host galaxy must be below this level to be obscured by scattering in the Milky Way.

As in the previous case study, we will start by deriving properties of the ISM local to the source of \RIII{} from optical observations of the host galaxy. \citet{marcote_repeating_2020} estimate the projected size of the emission region to be $1.5$\,kpc$^2$.  In a region of $1.5\times2$\,arcsec$^2$ surrounding the FRB localization, \citet{marcote_repeating_2020} measure an extinction-corrected H${\rm \alpha}$ luminosity of $L_{\mathrm{H}\alpha} \sim (2.0 \pm 0.1)\times10^{39}\,\mathrm{erg}\,\mathrm{s}^{-1}$, which they suggest is dominated by the star-forming clump. The H${\rm \alpha}$ line at the location of the FRB is measured by \citet{marcote_repeating_2020} to have a centroid and width of 6787.5\,\AA{} and 6.5\,\AA{} respectively. We will adopt the calibration precision of 0.8\,\AA{} as the uncertainty on both the line centroid and width.\footnote{\citet{marcote_repeating_2020} obtained these spectra with the GMOS instrument with R640 and measure sky lines with 7–8\,\AA{} widths, suggesting that the H$\alpha$ line is unresolved.}  Estimating the thermal velocity dispersion as $9.1\pm2.7$\,km\,s$^{-1}$, we calculate a non-thermal velocity dispersion of $290\pm50$\,km\,s$^{-1}$ from the H${\rm \alpha}$ line width measured by \citet{marcote_repeating_2020}. 

Through detailed HST observations, \citet{tendulkar_60_2020} found that this V-shaped emission region is composed of many clumps.  The one closest in projected distance from \RIII{} is at the vertex of the V, with a H${\rm \alpha}$ luminosity of $9\times10^{37}$\,erg\,s$^{-1}$ (detected after smoothing the original 30\,mas HST WFC3/F673N emission maps with a Gaussian kernel with FWHM 294\,mas; we take 20\% uncertainty on this measurement).  \citet{tendulkar_60_2020} measure a projected distance between this star-forming region, which has a size of 270\,mas or 190\,pc, and \RIII{} of 250$\pm$190\,pc.  We will adopt a path length through the region of $L=190\pm100$\,pc. We estimate the EM from the H${\rm \alpha}$ luminosity of the region according to equation (3) of \citet{tendulkar_host_2017} (see also \citet{reynolds_pulsar_1977}),
\begin{equation}
    \mathrm{EM}(\mathrm{H}\alpha) = 2.75\,\mathrm{pc}\,\mathrm{cm}^{-6} T_4^{0.9} \left( \frac{S(\mathrm{H}\alpha)_s}{\mathrm{Rayleigh}}\right)\,
\end{equation}
where $S(\mathrm{H}\alpha)_s$ is the H${\rm \alpha}$ surface density given in the source frame,
\begin{equation}
    S(\mathrm{H}\alpha)_s = (1+z)^4 \frac{F_{\mathrm{H}\alpha}}{\pi a b}\;.
\end{equation}
$F_{\mathrm{H}\alpha}$ is the H${\rm \alpha}$ flux, and $a$ and $b$ are the semi-major and semi-minor axes respectively.  For \RIII{}, we find $\mathrm{EM}=310\pm140\,\mathrm{pc}\,\mathrm{cm}^{-6}$.  This corresponds to electron densities of 0.2$\pm$0.10 and 0.6$\pm$0.3\,cm$^{-3}$ for filling factors of 0.1 and 1 (assuming $\zeta=2$ and $\epsilon=1$), and this region would contribute a DM in the observer frame between 10 and 170 to the FRB.  This is consistent with the host DM limit of $<70$ for \RIII{} \citep{marcote_repeating_2020}.

In the absence of any observed extragalactic scattering towards \RIII{}, we will use the scattering upper limit of $\tau < 1.6$\,ms at 350\,MHz \citep{chime_r3lowfreq_2020} to place an upper limit on the non-thermal velocity dispersion in the ISM local to \RIII{} predicted from radio-wave scattering. Assuming a distance between the lens and the FRB source of $d_\mathrm{lens,src} = 100\pm50$\,pc, we find $r_\mathrm{diff} > 3.2\times10^7$\,cm and $\mathrm{SM} < 0.8\,\mathrm{kpc}\,\mathrm{m}^{-20/3}$. From the SM and EM, we calculate an expected non-thermal velocity dispersion of $<480$\,km\,s$^{-1}$ for $f_f=0.1$ and $f_r=1$. Varying $f_f$ between 0.1 and 1, and $f_r$ between 0.5 and 1, we predict $\sigma_v < 620$\,km\,s$^{-1}$, similar to the velocity dispersion of $290\pm50$\,km\,s$^{-1}$ calculated from the observed H${\rm \alpha}$ line width. The energy in turbulence corresponding to both the constraint on scattering from the burst observations and the optical velocity dispersion are listed in Table \ref{tbl:results}.  In this case, the limit from scattering is consistent with the H${\rm \alpha}$ line width measured at the location of the FRB by \citet{marcote_repeating_2020}.

\subsection{\RI{}}\label{sec:R1}

\RI{} was the first FRB to be observed to repeat \citep{spitler_repeating_2016} and the first to be localized and associated with a host galaxy \citep{chatterjee_r1localization_2017}.  The host galaxy of \RI{} is a star-forming dwarf galaxy at a redshift $z=0.19273(8)$ \citep{tendulkar_host_2017}.  VLBI observations revealed a persistent radio source close to the FRB position \citep{marcote_repeating_2017}.  High-resolution observations of the host galaxy reveal a star-forming region that encompasses both the FRB location and the associated persistent radio source \citep{kokubo_Halpha_2017,bassa_121102_2017}.

Again, we start by extracting properties of the ISM local to the FRB source from optical observations. \citet{bassa_121102_2017} measure the projected half-light radius of the star-forming region in the F101W filter of their HST observations to be $0.20\pm0.01$\,arcsecond, giving a diameter of the star-forming region of $1.32\pm0.07$\,kpc (using an angular diameter distance of $682.2\pm0.2$\,Mpc). This is consistent with the upper limit on the FWHM of the H${\rm \alpha}$-emitting region, $<0.57$\,arcsecond, placed by  \citet{kokubo_Halpha_2017} using Subaru/Kyoto H${\rm \alpha}$ images.  Within the star forming region, \citet{kokubo_Halpha_2017} measure an H${\rm \alpha}$ flux consistent with that reported by \citet{tendulkar_host_2017} for the entire galaxy, indicating that the H${\rm \alpha}$ flux measured by \citet{tendulkar_host_2017}, $2.6\times10^{-16}\,\mathrm{erg}\,\mathrm{cm}^{-2}\,\mathrm{s}^{-1}$, is dominated by the star-forming region.  From the H${\rm \alpha}$ flux from \citet{tendulkar_host_2017} and the area of the star-forming region measured by \citet{bassa_121102_2017}, we calculate $\mathrm{EM} = 510\pm160$\,pc\,cm$^{-6}$. Assuming a path length through the region equal to the region's diameter, this corresponds to an electron density of $0.097\pm0.016$ or $0.31\pm0.05$\,cm$^{-3}$ for filling factors of 0.1 or 1 (assuming $\zeta=2$ and $\epsilon=1$).  If the FRB path length through the region is equivalent to the half-light radius of the region, this would contribute a DM between 50 and 400 to the FRB, consistent with the analysis of \citet{tendulkar_host_2017} based on the H$\alpha$ emission of the host galaxy and the 340\,pc\,cm$^{-3}$ extragalactic \citep[IGM + host galaxy + local environment,][]{tendulkar_host_2017} DM of \RI{}. \citet{tendulkar_host_2017} measure a width of the H${\rm \alpha}$ line of $2.02\pm0.03$\,\AA{}.\footnote{\citet{tendulkar_host_2017} measure the width of the H$\alpha$ line with the Gemini/GMOS R400 grating (resolution 4.66\,\AA{}) and therefore have underresolved the H$\alpha$ line.  The true width of the line is likely much smaller, and the line-width measured from high-resolution spectra with a 1200 lines/mm grating in the Keck/LRIS instrument is $<1$\,\AA{} (Chen et al.\ in prep.).} From this width, we measure a velocity dispersion in the star-forming region of $92.2\pm1.4$\,km\,s$^{-1}$.  Subtracting a thermal contribution of $9.1\pm2.7$\,km\,s$^{-1}$ in quadrature, we estimate a non-thermal velocity dispersion of $91.7\pm1.6$\,km\,s$^{-1}$.

As was the case for \RIII{}, there is no evidence of extragalactic scattering towards \RI{} from radio observations, but we can still place an upper limit on the scattering measure and the predicted non-thermal velocity dispersion in the local ISM. The best constraints on scattering of \RI{} come from observations with the CHIME/FRB instrument, which yield an upper limit on the scattering timescale of 9.6\,ms at 500\,MHz \citep{chime_r1detection_2019}.  Assuming the source and scattering screen are separated by a distance $d_\mathrm{lens,src} = (0.5 \pm 0.25)\,\mathrm{FWHM}_\mathrm{SFR}$, we limit the diffractive scale and the SM to be $r_\mathrm{diff} > 2.9\times10^7$\,cm and $\mathrm{SM}< 4.3$\,kpc\,m$^{-20/3}$.

Combining the scattering and emission measures, we find, for $f_r=1$ and $f_f=0.1$, an expected velocity dispersion of $\sigma_v < 1100$\,km\,s$^{-1}$. For $f_R=0.5$ and $f_f=0.1$, $\sigma_v < 1600$\,km\,s$^{-1}$. This constraint is consistent with the optical measurement of $91.7\pm1.6$\,km\,s$^{-1}$, but the SM is sensitive to only much larger velocity dispersions than that measured from the emission line width.  This suggests that, for \RI{}, current radio burst observations do not provide strong constraints on the turbulence in the plasma.  A burst detection at lower frequency, where radio-wave scattering is stronger, may provide a stronger constraint. However, \citet{spitler_5GHz_2018} measure, at 5\,GHz, a scintillation bandwidth of $6.4\pm1.6$\,MHz consistent with expectations of Galactic scattering.  At 300\,MHz, this will result in a scattering timescale of $\sim 4$\,ms, which would mask extragalactic scattering below this level.  The energies calculated from both the velocity dispersion constraint from scattering and the non-thermal velocity dispersion calculated from the H${\rm \alpha}$ line width are shown in Table \ref{tbl:results}.

With only an upper limit on the SM, we can still place a lower limit on the outer scale if we assert that the scattering is probing the same turbulent cascade as the gas seen in emission and the scales probed by radio-wave scattering and the H${\rm \alpha}$ line width are connected by a Kolmogorov turbulence power spectrum.  In this case, we find $l_o > 4.4\times10^{15}$\,m ($> 0.014\,$pc), consistent with our expectations of an outer scale on the order to $100\,$pc (approximately the size of the star-forming region).

\section{Discussion and Conclusions}\label{sec:conclusions}

\subsection{Summary of results}

In this work:
\begin{enumerate}
\item We have presented a formalism for combining FRB SMs with optical followup of FRB host galaxies to trace density and velocity fluctuations in the ISM of these host galaxies on multiple scales. As a method of comparison, we use the turbulent energy inferred from these observables. The turbulent energy is related to the SM (sensitive to density fluctuations on the Fresnel scale, $\sim10^8$\,m) and non-thermal velocity dispersion (sensitive to velocity fluctuations on the outer scale, $\sim10^{18}$\,m) by equations \eqref{eqn:turb_energy} and \eqref{eqn:turb_energy_sigmav} respectively. By inferring the turbulent energy on these vastly different scales, we can constrain the nature of the turbulent cascade in the ISM of galaxies besides the Milky Way, allowing us to study multiple instances of density-fluctuation spectra. At the same time, this methodology provides insight into the local environments of FRB sources and the origin of the observed scattering.
\item We have shown that upcoming FRB surveys will be able to use the presence of Galactic scintillation to determine the location of FRB scattering screens. For example, over $>15\,\%$ of the visible sky, the DSA-110 will be able to determine whether extragalactic scattering is originating from within 50\,kpc of the FRB source of $z\leq0.2$ FRBs scattered on 1\,ms timescales at 1.405\,GHz.  Surveys at lower frequencies and with wider bandwidths will be able to make this determination over a much larger part of the sky; for example $>99\,\%$ of the sky for a CHORD observation of a similarly scattered FRB between 300 and 700\,MHz.
\item We have presented three case studies to demonstrate the power of this technique.  In all cases, we compare measurements or limits on extragalactic scattering to properties of HII regions near the FRB sources in their host galaxies.
\begin{enumerate}
    \item \ASKAPFRB{}, with an extragalactic scattering timescale of $3.3\pm0.2$\,ms at 1200\,MHz that most likely originates from the host galaxy or source local environment \citep{chittidi_dissecting_2020} although a search for Galactic scintillation was inconclusive \citep{day_high_2020}.  From the scattering timescale, we infer a turbulent velocity dispersion $\sim 4$ orders of magnitude higher than that measured from the H${\rm \beta}$ line width.  Such a large discrepancy cannot be explained by varying the outer and inner scales, but a decrease in the index of the density fluctuation power spectrum to $\beta \sim 3.2$ can reconcile this difference.
    \item \RIII{}, with a limit on extragalactic scattering of $>1.6\,$ms at 350\,MHz, set by Galactic scattering towards \RIII{}.  The upper limit on the turbulent energy inferred from the scattering timescale is approximately twice the measurement from the H${\rm \alpha}$ line width. These measurements are consistent with a Kolmogorov turbulent cascade over the ten orders of magnitude between the scales probed by the two processes.
    \item \RI{}, with a limit on the extragalactic scattering of $<9.6\,$ms at 500\,MHz.  The upper limit on the turbulent energy inferred from the scattering timescale is more than two orders of magnitude larger than the measurement from the H${\rm \alpha}$ line width, suggesting that the scattering timescale constraints are insufficient for this type of study in \RI{}.
\end{enumerate}
\end{enumerate}

\subsection{Improving constraints on the density fluctuation power-spectrum}

The energy derived from the scattering timescale (equation \eqref{eqn:sigmav_SM_1}) depends strongly on the form of the turbulent power spectrum that we assume, including the power-law index, the inner scale ($E\propto l_i^{1/3}$ for Kolmogorov turbulence), and the outer scale ($E \propto l_o^{2/3}$, which we associate with the size of the nearby star-forming regions in our case studies), as well as properties of the scattering medium, including the electron density ($E \propto \langle n_e \rangle^2$, which we have constrained through the EM) and the line-of-sight separation between the source and the scattering material.  We also note that we have considered solely isotropic turbulence in this work; anisotropic turbulence could also be considered with modification of the relations between the observables (scattering timescale and non-thermal velocity dispersion) and the amplitude of the density and velocity fluctuations.

There are a number of factors here that can be better constrained. To determine the EM of the H$\alpha$-emitting regions from the H$\alpha$ flux (and derive their electron densities), we need their spatial sizes. Strong constraints on these sizes from high-spatial resolution optical observations of FRB host galaxies are therefore important for constraining the electron density. 

The properties of the H$\alpha$-emitting gas can also be probed directly from the observed emission lines.  High-resolution spectral line observations are critical to constraining the properties of this gas. In particular, the SII (6716\,\AA{} or 6731\,\AA{}) to H$\alpha$ line-width ratio provides a direct decomposition of the thermal and non-thermal components to the velocity, due to the high mass ratio of these two elements (32 to 1) \citep[e.g.][]{reynolds_SII_1985}. The SII to H$\alpha$ line flux ratio is also a useful diagnostic of gas conditions, including the ionization state. The electron densities of $\sim 0.1$\,cm$^{-3}$ inferred from the EMs in Section \ref{sec:case_studies}, the high SII 6716\,\AA{} to H$\alpha$ flux ratio of $0.24\pm0.07$ in the local environment of \RIII{} \citep{marcote_repeating_2020}, and the non-detection of a point source of H$\alpha$ emission near \RIII{} \citep{tendulkar_60_2020} all suggest that the H$\alpha$ emission in the local environment of \RIII{} is dominated by diffuse ionized gas like the Milky Way WIM (diffuse ionized gas present throughout the Milky Way with temperatures $\sim$6000 to 10000\,K and generally lower ionization states than classical HII regions), or the diffuse ionized gas (DIG) of extragalactic galaxies.\footnote{See \citet{hdb+09} for a review on the WIM and DIG. The line ratios of the SII forbidden lines to H$\alpha$ are many times larger in the WIM than in HII regions \citep{reynolds_SII_1985}.} A different picture emerges for \RI{}; the low SII 6716\,\AA{} to H$\alpha$ line flux ratio (0.06$\pm$0.01 after correcting for extinction) observed by \citet{tendulkar_host_2017} from the host galaxy of \RI{} is more consistent with an HII region than the Milky Way WIM. The flux ratio between the 6716\,\AA{} and 6731\,\AA{} components of the SII doublet can also be used to constrain the electron density independently when it is above 37\,cm$^{-3}$ \citep{osterbrock_2006}.  The electron densities inferred from the EMs in our case studies are orders of magnitude below this value.  The conclusion that the H$\alpha$ emitting gas in these regions of interest is too tenuous to measure the density using the SII line ratio is supported by the SII (6716\,\AA{}/6731\,\AA{}) line ratios for \RI{} (1.7$\pm$0.5; spectra from \citep[1.7$\pm$0.5, spectra from][]{tendulkar_host_2017} and \RIII{} \citep[2.0$\pm$0.3, spectra from][]{marcote_repeating_2020}, both larger than the $\lesssim1.4$ ratio needed for measuring the electron density from the SII line ratio. These case studies show that high-spectral-resolution  observations of the SII and H$\alpha$ emission lines is critical for studying the gas near an FRB source, and can probe the ionization state, temperature and density of the gas.  This is especially powerful when combined with a size estimate of the H$\alpha$ emitting region from imaging; the temperature and size of the H$\alpha$ region reduce the number of variable parameters when inferring the electron density from the H$\alpha$ flux.

Our results also depend on assumptions for the inner and outer scale, but, as we see in our study of \ASKAPFRB{}, the most important uncertainty in the turbulent power spectrum is the scaling index $\beta$. Thus, one of the best ways to improve this comparison between density and velocity fluctuations on different scales is to add tracers that can constrain $\beta$.  In the Milky Way, the fringe tilt in pulsar dynamic spectra, refractive scintillation, and dispersion measure (DM) fluctuations are used as probes of intermediate scales of $\sim10^{10}$ to $10^{14}$\,m.  Refractive scintillation, variation in the apparent flux of a source due to changes in the angular broadening scale, is not a good tracer of FRB local environments; angular broadening due to scattering in plasma close to the source is suppressed compared to that due to scattering close to the observer, as discussed in Section \ref{sec:vlbi}. \citet{yang_dispersion_2017} consider possible sources of DM variations in repeating FRB sources, and predict contributions of $\sim 2\times10^{-5}$\,pc\,cm$^{-3}$\,yr$^{-1}$ due to turbulent fluctuations in interstellar plasma with density fluctuations of $\sim 0.1$\,cm$^{-3}$ (similar to the densities inferred from the EMs of the three FRBs presented in Section \ref{sec:case_studies}, which range from $\sim 0.1$ to $\sim 0.9$\,cm$^{-3}$ for filling factors between $0.1$ and $1$).  This is on the order of the best constraints for pulsar DM variations \citep[e.g.][]{you_dispersion_2007}.  However,  \citet{yang_dispersion_2017} find that DM variations due to expansion of a supernova remnant or HII region, or the proper motion of the FRB source with respect to an HII region, are much larger and, if present, will mask any smaller contributions due to density fluctuations in the plasma. The complex spectrotemporal structure of fast radio bursts further complicates detecting small-scale DM variations.  Typically, FRB DM uncertainties are on the order of $\sim 0.1$\,pc\,cm$^{-3}$, although \citet{nimmo_microsecond_2020} measure the DM of \RIII{} to 0.006\,pc\,cm$^{-3}$ precision. The DM of \RI{} does show variation over time \citep{hessels_frb_2019} but the fluctuations expected from density variations have not been observed;  \citet{hilmarsson_rotation_2020} find a steady increase in the DM of \RI{} over a period of two years. \RIII{} shows no noticeable DM variations \citep[e.g.][]{pastor_chromatic_2020}. 

Dynamic spectra are difficult to construct for repeating FRBs due to their low duty cycles.  As repeat bursts tend to be temporally clustered \citep{oppermann_nonpoissonian_2018, chime_periodic_2020}, observations of repeating FRB sources with highly sensitive radio instruments, such as FAST, may provide sufficient temporal sampling to construct a dynamic spectrum. However, it is possible that much of the spectrotemporal structure, including the downward marching `sad trombone' effect seen in repeating FRB dynamic spectra \citep{hessels_frb_2019} is intrinsic; this intrinsic structure must be disentangled from any propagation-induced scintillation to use the dynamic spectrum as a probe of local turbulence. As our understanding of intrinsic FRB spectral structure advances, dynamic spectra and DM variations of repeating FRB sources may prove valuable ways of studying FRB environments; however, these methods do not take full advantage of the capability of instruments like ASKAP, the DSA and CHORD to localize non-repeating sources of FRBs, which may be a distinct class from FRB repeaters.

Multiple tracers at similar scales can be used to constrain the index of the fluctuation power spectrum (even if only at one end of the spectrum), adding some confidence to an extrapolation over orders of magnitude. If some optical spectral lines are resolved, additional information on the turbulent velocity power spectrum is present in the shape of the spectral line.  This arises from Doppler shifting of the line by the turbulent velocity fluctuations.  \citet{lazarian_2000,lazarian_2004,lazarian_2006} have developed the formalism for relating the line profile to the turbulent power spectrum, and, using this technique, the power-law spectral index and the outer scale of turbulence in neutral hydrogen have been measured at high latitudes in the Milky Way \citep{chepurnov_2010} and in the Small Magellanic Cloud \citep{chepurnov_2015}.

At the opposite end of the spectrum, pulsar scattering indices from multi-band or wide-band scattering timescale measurements can be used to constrain the index of the fluctuation power spectrum at small spatial scales. For isotropic turbulence following a power-law spectrum, $\tau \propto \nu^{-\alpha}$ where $\alpha = 2\beta/(\beta-2)$; however low values of $\alpha$ can be inferred if anisotropic scattering is not properly modelled \citep{geyer_scattering_2017, oswald_thousand_2021}. \citet{day_high_2020} apply this technique to four FRBs discovered and localized by ASKAP, including \ASKAPFRB{}, as discussed in Section \ref{sec:case_studies}.

FRBs present the only means of probing density fluctuations in extragalatic ISM on $\ll$pc scales. The techniques we have developed will enable the use of FRBs to compare density-fluctuation measurements on scales across the putative turbulent cascades, extending these studies beyond the Milky Way for the first time. These considerations are also central in addressing the complementary question of the origin of observed FRB scattering. Even within the Milky Way, the physical origins of interstellar scattering remain inscrutable to this day, as the phenomenology of scattering grows richer. As FRB samples expand and localization accuracies improve, we anticipate that some of the most exciting discoveries about plasma along FRB sightlines may be revealed by scattering measurements. 

\begin{acknowledgements}

We thank Phil Hopkins for many useful discussions and comments during this work. This research was supported by the National Science Foundation under grant AST-1836018.

\end{acknowledgements}

\bibliography{main}{}
\bibliographystyle{aasjournal}

\appendix
\section{Derivation of the scatter broadening angle on the extragalactic screen}\label{sec:theta}

\begin{figure}
    \centering
    \includegraphics[width=0.3\textwidth]{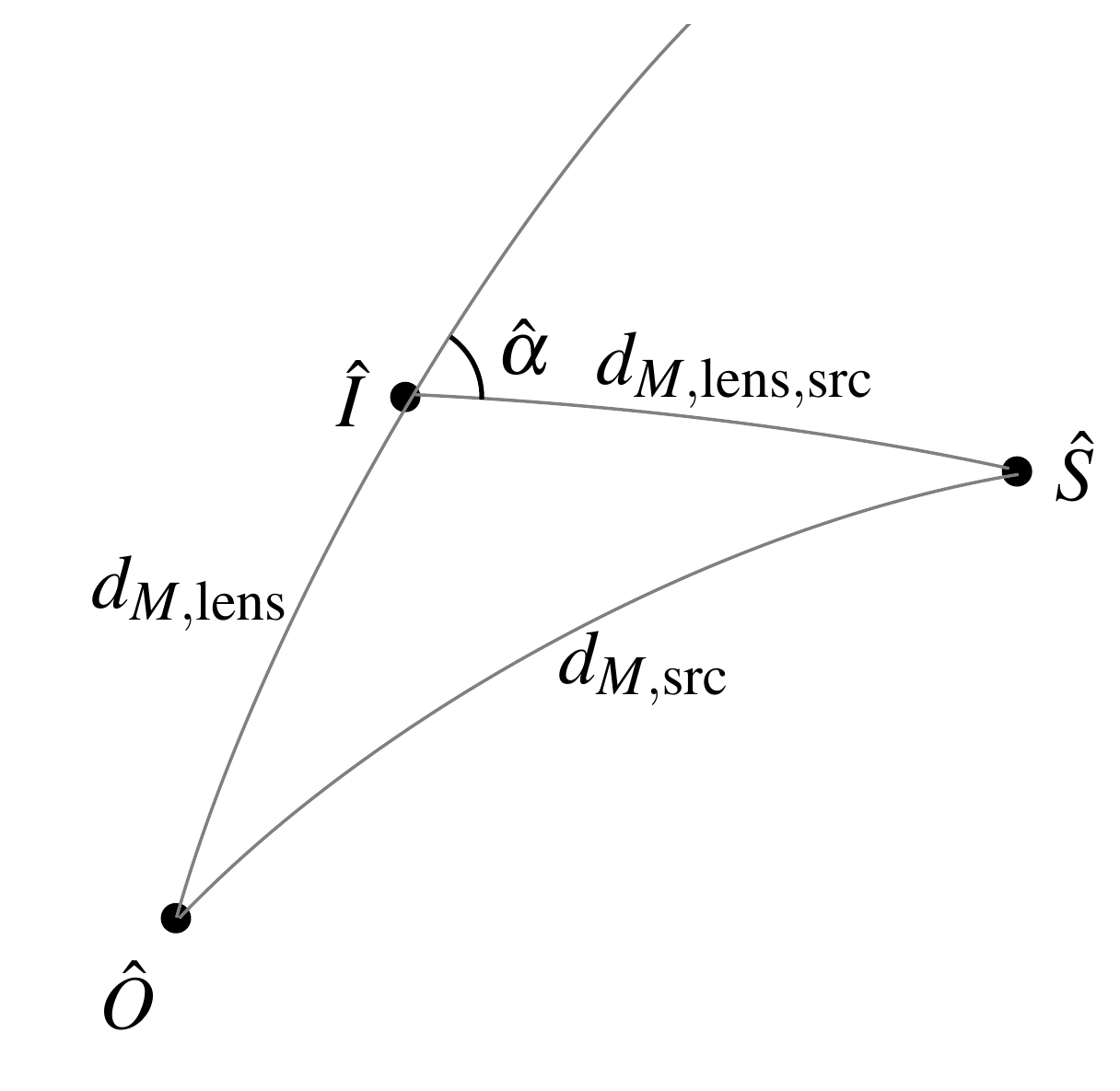}
    \caption{The comoving geometry.  $d_{M,\mathrm{src}}$, $d_{M,\mathrm{lens}}$ and $d_{M,\mathrm{lens,src}}$ are the comoving distances between the observer and the source, the observer and the lens, and the lens and the source respectively.  This figure is similar to  \protect\citet{schneider_gravitational_1992}, their Fig.\ 4.1, but specialized to a flat universe.}
    \label{fig:geometry_comoving}
\end{figure}

To derive the scattering broadening angle, $\theta_\mathrm{scat}$, for scattering at an extragalactic scattering screen, we work along the lines of \citet[][Chapter 4.6]{schneider_gravitational_1992}.  For simplicity, we will assume a flat universe; \citet{schneider_gravitational_1992} show that the same relation holds in a positively-curved universe.  Our setup is shown in Fig.\ \ref{fig:geometry_comoving}.  We start with a transformation to comoving coordinates and conformal time, choosing a conformal factor at the observer of $a(z_\mathrm{obs})=1$, and write the geometric time delay imparted by lensing at a point $\hat{I}$ of radiation emitted by a point source at $\hat{S}$ as seen by an observer at $\hat{O}$. \begin{equation}
    c \Delta t_\mathrm{geom} = \Delta \eta_\mathrm{geom} = \frac{d_{M,\mathrm{lens,src}} d_{M,\mathrm{lens}}}{2 d_{M,\mathrm{src}}} \hat{\alpha}^2\;.
\end{equation}
We use the relation between comoving distances ($d_M$) and angular diameter distance measured from $z_1$ to $z_2$ (for a flat universe),
\begin{equation}\label{eqn:angdiam_comoving}
    d(z_1, z_2) = \frac{1}{1+z_2}(d_{M,2} - d_{M,1})\;,
\end{equation}
and write
\begin{equation}\label{eqn:tgeom}
    c \Delta t_\mathrm{geom} = \frac{d_\mathrm{lens,src} d_\mathrm{lens}}{2\,d_\mathrm{src}} (1+z_\mathrm{lens}) \hat{\alpha}^2\;.
\end{equation}
$t_\mathrm{geom}$ is the geometric time delay in conformal time. The geometric part of the phase difference is then
\begin{equation}
    \Delta \phi_\mathrm{geom} = c \Delta t_\mathrm{geom} \Omega
\end{equation}
where $\Omega$ is the circular frequency in the conformal space.  $\Omega$ is related to the frequency in the observer's frame, $\omega$ by $\omega = a(z_\mathrm{obs}) \Omega$. Our choice of conformal factor has thus ensured that $\omega = \Omega$, and we can write the phase imparted as
\begin{align}
    \Delta \phi_\mathrm{geom} &= c \Delta t_\mathrm{geom} \frac{2\pi}{\lambda_\mathrm{obs}}\\
    &= \frac{2\pi}{\lambda_\mathrm{obs}} \frac{d_\mathrm{lens,src} d_\mathrm{lens}}{2d_\mathrm{src}} (1+z_d) \hat{\alpha}^2;.
\end{align}

\begin{figure}
    \centering
    \includegraphics[width=0.4\textwidth]{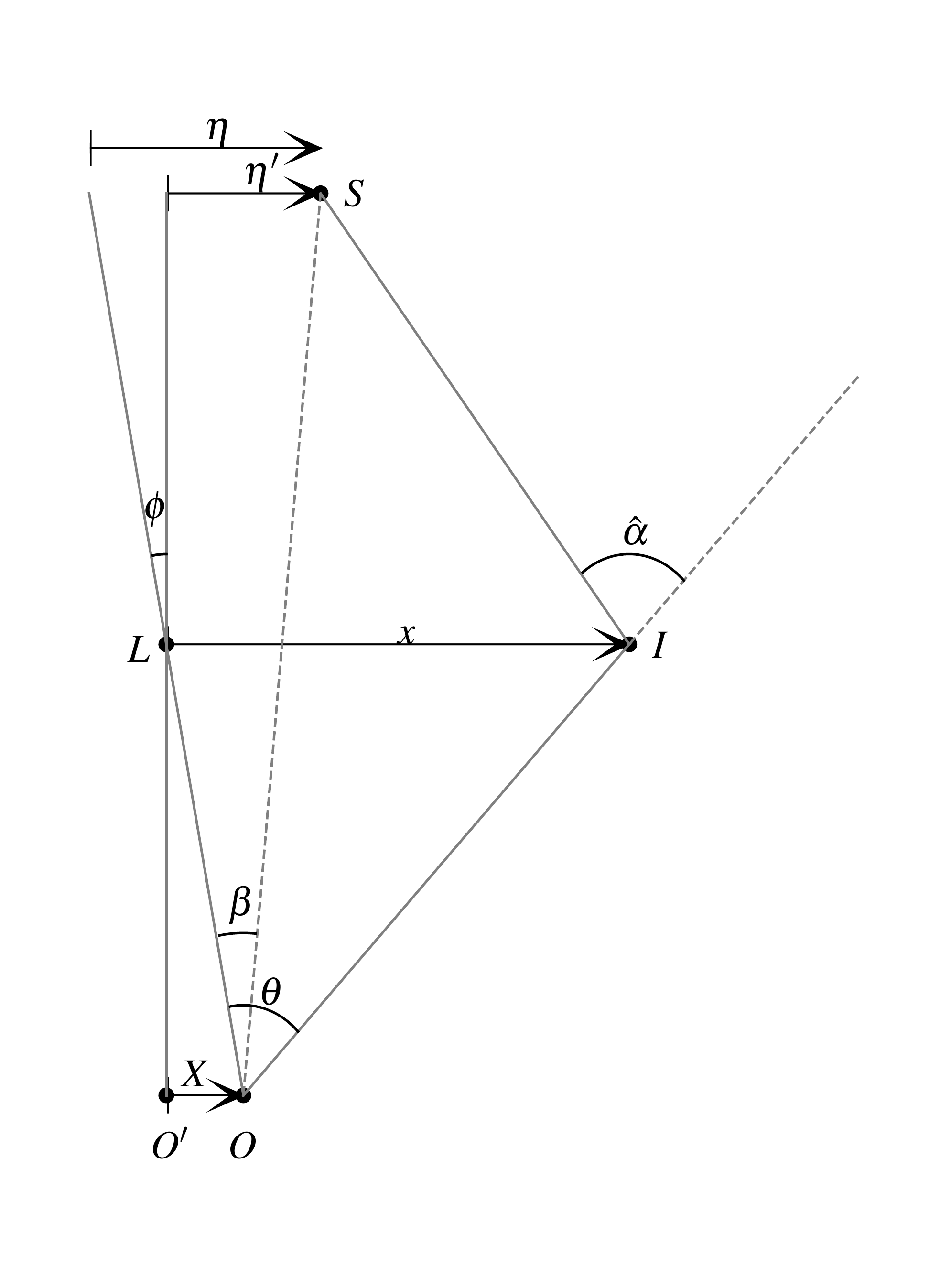}
    \caption{The lensing geometry. $\eta$, $x$ and $X$ are spatial coordinates on the source, lens, and observer planes respectively. $S$, $I$, and $O$ are the locations of the source, image, and observer, while $O^\prime$ and $L$ define our reference on all plane.}
    \label{fig:geometry_general}
\end{figure}

In this derivation, we have used a special reference line which intersects $\hat{O}$ and $\hat{I}$. We need to generalize this to any reference line, as shown in Fig.\, \ref{fig:geometry_general}, to obtain the scattered wavefield at any point on the observer plane through the Fresnel-Kirchoff integral,
\begin{align}
    u(\mathbf{X}) &= \frac{e^{-i\pi/2}}{2\pi r_F^2} \int d^2\mathbf{x} \exp\left[ i \Delta \phi(\mathbf{x}, \mathbf{X})\right]\\
    &= \frac{e^{-i\pi/2}}{2\pi r_F^2} \int d^2\mathbf{x} \exp\left[ \frac{2 \pi}{\lambda_\mathrm{obs}} i c \tau(\mathbf{x}, \mathbf{X})\right]\;,
\end{align}
where $\tau(\mathbf{x}, \mathbf{X}) = \Delta t_\mathrm{geom}(\mathbf{x}, \mathbf{X}) + \frac{\phi(\mathbf{x})}{2\pi\nu}$. Here, $\mathbf{x}$, $\mathbf{X}$ and $\bm{\eta}$ are spatial coordinates at the lens, observer, and source planes, respectively, and $\phi(\mathbf{x})$ is the part of the phase independent of the observer position, which includes the phase imparted by the lens. The Fresnel scale, generalized to this cosmology, is given in equation \eqref{eqn:fresnel}.  We will ignore the delay imparted by travel through the lens, assuming that its variation with $\mathbf{x}$ is much smaller than the variation of the geometric delay with $\mathbf{x}$.

As shown in Fig.\ \ref{fig:geometry_general}, we have the following geometric relations:
\begin{align}
    \hat{\bm{\alpha}} d_\mathrm{lens,src} &= (\bm{\theta}-\bm{\beta}) d_\mathrm{src}\;,\label{eqn:alpha_theta}\\
    \bm{\eta} &= \bm{\beta} d_\mathrm{src}\;,\\
    \bm{\eta}' &= \bm{\eta} - \bm{\phi} d_\mathrm{lens,src}\;,\\
    \mathbf{x} &= \bm{\theta} d_\mathrm{lens}\;\mathrm{and}\\ 
    \mathbf{X} &= \bm{\phi} d_\mathrm{lens,obs} = \bm{\phi} d_\mathrm{lens} (1+z_\mathrm{lens})\;.
\end{align}
In the last line, we've used $d_\mathrm{lens,obs}$ to denote the angular diameter distance from the lens to the observer, and equation \eqref{eqn:angdiam_comoving} to write $d_\mathrm{lens,obs} = d_\mathrm{obs,lens} \frac{1+z_\mathrm{lens}}{1+z_\mathrm{obs}} = d_\mathrm{lens} (1+z_\mathrm{lens})$.

Putting these all together with equation \eqref{eqn:tgeom}, we can write the geometric time delay in terms of $\mathbf{X}$, $\mathbf{x}$ and $\bm{\eta}'$:
\begin{equation}\label{eqn:tgeom2}
    c \Delta t_\mathrm{geom} = \frac{d_\mathrm{src}}{2 d_\mathrm{lens} d_\mathrm{lens,src}}(1+z_\mathrm{lens}) \left( \mathbf{x} - \frac{d_\mathrm{lens}}{d_\mathrm{src}}\bm{\eta}' - \frac{d_\mathrm{lens,src}}{d_\mathrm{src}}\frac{1}{1+z_\mathrm{lens}}\mathbf{X}\right)^2\;.
\end{equation}
For a point source, we can choose $\bm{\eta}'=0$ and simplify this expression.  We note that this differs from equation (3) of \citet{macquart_temporal_2013}, who have omitted the $(1+z_\mathrm{lens})^{-1}$ factor in the $\mathbf{X}$ term, we believe in error.  However, this does not effect their derived scattering timescale, and therefore the bulk of their conclusions, which focus on redshift dependencies of scattering in host galaxies, the IGM, intervening galaxies, and the Milky Way.

Finally, we can follow the discussion in Section 2.1 of \citet{macquart_temporal_2013} to obtain the HWHM scattering angle from the wavefield from the average visibility
\begin{align}
    \langle V(\mathbf{r}) \rangle &= \langle u(\mathbf{X}'+\mathbf{r}) u^*(\mathbf{X}') \rangle \\
    &= \frac{I_0}{(2\pi r_F^2)^2} \int d^2\mathbf{x} d^2\mathbf{x}' \exp\left[ \frac{i}{2r_F^2}\left(\mathbf{x}-\frac{d_{\mathrm{lens,src}}}{d_\mathrm{src} (1+z_\mathrm{lens})}(\mathbf{X}'+\mathbf{r})\right)^2 - \frac{i}{2r_F^2} \left( \mathbf{x}'-\frac{d_\mathrm{lens,src}}{d_\mathrm{src} (1+z_\mathrm{lens})}\mathbf{X}'\right)^2\right]\label{eqn:vis}\\
    &= I_0 \exp \left[-\frac{1}{2}D_\phi\left(\frac{d_\mathrm{lens,src}}{d_\mathrm{src} (1+z_\mathrm{lens})}\mathbf{r}\right)\right]\;,
\end{align}
where $D_\phi(\mathbf{r})$ is the structure function describing the phase fluctuations on the scattering screen,
\begin{equation}
D_\phi(\mathbf{r}) = <[\phi(\mathbf{r}+\mathbf{r}')-\phi(\mathbf{r}')]^2>\,.
\end{equation}
For a power-law spectrum of isotropic fluctuations,
\begin{equation}
    D_\phi(r) = \left(\frac{r}{r_\mathrm{diff}}\right)^{\beta-2}\;,
\end{equation}
where the diffractive scale $r_\mathrm{diff}$ is given by equation \eqref{eqn:rdiff} and $\beta$ is the same index as in equations \eqref{eqn:P_n} and \eqref{eqn:P_v}. 

The image on the sky is related to the visibility by the van Cittert-Zernicke theorem,
\begin{equation}
    I(l, m) = \int \mathrm{d}u \int \mathrm{d}v V(u, v) \exp(2\pi i (ul + vm))\;,
\end{equation}
where $\mathbf{r} = \lambda (\mathbf{u}+\mathbf{v})$. For the choice $\beta=2$, $V(u,v)$ is a Gaussian function, and
\begin{equation}
    I(l,m) \propto \exp \left( -2\pi^2 \left(\frac{d_\mathrm{src} (1+z_\mathrm{lens})}{d_\mathrm{lens,src} \lambda}\right)^2 r_\mathrm{diff}^2 \theta^2 \right)\;,
\end{equation}
where $\theta^2 = l^2 + m^2$.  Then the half-width half-max scattering angle is
\begin{equation}\label{eqn:theta}
    \theta_\mathrm{HWHM} = \sqrt{2\ln(2)} \frac{\lambda}{2\pi} \frac{d_\mathrm{lens,src}}{d_\mathrm{src} (1+z_\mathrm{lens})}\frac{1}{r_\mathrm{diff}}\;.
\end{equation}
As \citet{macquart_temporal_2013} note, different values of $\beta$ change the numerical prefactor in the scattering angle.

We find that angular broadening is largest for scattering at low redshifts. This differs from equation (13) of \citet{macquart_temporal_2013} due to the difference in equation \eqref{eqn:tgeom2}.  Finally, we can use the relationship between the delay and diffractive scale (equation \eqref{eqn:rdiff}):
\begin{align}
    \tau &= \frac{\lambda}{2\pi c}\left(\frac{r_F}{r_\mathrm{diff}}\right)^2\\
    &= \frac{d_\mathrm{lens} d_\mathrm{src}}{c d_\mathrm{lens,src}} \theta_\mathrm{scatt}^2 (1+z_\mathrm{lens})\;,\label{eqn:tau_theta}
\end{align}
which again differs from equation (15) of \citet{macquart_temporal_2013}.

From equation \eqref{eqn:theta}, we see that, if $r_\mathrm{diff}$ in the host galaxies does not vary with redshift, then angular broadening due to scattering within the host galaxies of FRBs is $\propto d_\mathrm{src}^{-1} (1+z_\mathrm{src})^{-1}$. $d_\mathrm{src} (1+z_\mathrm{src})$ is a monotonically increasing function, and so angular broadening is much more likely to be detected from nearby FRBs.  In general, equation \eqref{eqn:theta} shows that angular broadening is dominated by scattering material close to the observer, where $d_\mathrm{lens,src}$ is large and $(1+z_\mathrm{lens})$ is small.

While the full analysis of the visibility in equation \eqref{eqn:vis} is necessary to determine how different forms of the turbulent structure function (for example, different power-law indices $\beta$) affect the relation between the $1/e$ scattering timescale and the HWHM of the scatter-broadened image, the redshift dependence in the relationship between the delay and broadening angle in equation \eqref{eqn:tau_theta} is easily seen from equations \eqref{eqn:tgeom2} and \eqref{eqn:alpha_theta}.

\section{The fraction of FRBs for which the scattering material location can be constrained from Galactic scintillation}\label{sec:appendixB}

In this appendix, we estimate the chances of observing FRBs that are both sufficiently close ($z<0.2$) for detailed optical follow-up of the host galaxies, and seen through a portion of the Milky Way scattering screen with sufficient resolution to effectively constrain a source-screen distance. We specifically consider cases where the source-screen distance can be constrained to $<50$\,kpc for an FRB that is scattered in the host galaxy to 1\,ms at 1.405\,GHz. We include the DSA-110, ASKAP CRAFT, CHORD, and DSA-2000 surveys in our analysis. 

We make use of the package frbpoppy \citep{gardenier_frbpoppycode_2019} to determine the redshift distribution of FRBs detected by these surveys.  The frbpoppy package is a population synthesis tool that incorporates observational biases and selection effects in order to consistently extract key properties of the intrinsic FRB population from the observed population; it has been applied to samples of both one-off \citep{gardenier_frbpoppy_2019} and repeating FRBs \citep{gardenier_frbpoppy_2021}.  We instead assume characteristics for the intrinsic FRB population in order to estimate the fraction of observed FRBs at low redshift ($z<0.2$).  We generate a one-off population of $10^4$ FRBs with a uniform direction distribution and a redshift distribution that tracks stellar mass density.  We adopt defaults for other characteristics of the population: We include DM contributions from the Milky Way (NE2001 model), IGM \citep[Ioka model with slope of 950;][]{ioka_cosmic_2003, inoue_probing_2004} and host (truncated Gaussian distribution with mean 100\,pc\,cm$^{-3}$ and standard deviation 200\,pc\,cm$^{-3}$). We assume a Gaussian distribution of spectral indices with a mean of $-1.4$ and standard deviation of 1 and a flat luminosity function between $10^{40}$ and $10^{45}$\,erg\,s$^{-1}$, and adopt a uniform distribution of intrinsic pulse widths between 0 and 10\,ms. We do not include the effects of scattering and scintillation in our rate estimates.

In Table \ref{tbl:surveys}, we show the results for the DSA-110 survey, the ASKAP CRAFT project \citep{macquart_craft_2010}, CHORD \citep{vanderlinde_chord_2019}, and the DSA-2000 \citep{hallinan_dsa_2019}. ASKAP CRAFT is a commensal survey with a maximum of 336\,MHz bandwidth with 1\,MHz channels between 700 and 1500\,MHz. We considered two possible setups, a low frequency band centered at 850\,MHz and a high-frequency band at 1350\,MHz. CHORD, currently in design stages, will observe from 300 to 1500\,MHz simultaneously; we have considered two bands of 400\,MHz at either end of this range, as well as the full frequency range.  The DSA-2000, also in design stages, will observe from 700 to 2000\,MHz simultaneously.  Again, we consider two bands of 400\,MHz as well as the full frequency range.  In all cases, we have scaled the fiducial extragalactic scattering delay from 1\,ms at 1405\,MHz using $\tau \propto \nu^{-4.4}$.  For ASKAP, we use a maximum declination of 30\,deg and for CHORD and the DSA-2000, minimum declinations of $-10$\,deg and $-30$\,deg respectively.  We use parameters for CHORD as included in the list of surveys in frbpoppy (version 2.1.0).  While the coherent ASKAP CRAFT survey is not included in the current version of frbpoppy, the incoherent ASKAP survey is, and we use these parameters for ASKAP CRAFT with the exception of the gain, which we multiply by 6 to account for the factor of $\sqrt{N}$ of increased sensitivity in a coherent search.  Parameters for the DSA-10 FRB survey are also included in frbpoppy, and we adopt these for the DSA-110 and DSA-2000 surveys with the following exceptions: 1.\ A minimum declination of $-30$\,deg.\ 2.\ Receiver temperatures of 30 and 25\,K for DSA-110 and DSA-2000 respectively.\ 3.\ Gains of 0.4 and 10.0\,K\,Jy$^{-1}$ for DSA-110 (assuming 85 dishes included in a coherent search) and DSA-2000 (assuming a coherent search with all 2000 dishes) respectively.\ 4.\ Sampling time of 1.048\,ms.

With these survey parameters, the DSA-110 and ASKAP CRAFT surveys are predicted to detect 0.0001\,\% of the intrinsic FRB population, while the DSA-2000 and CHORD surveys, with much greater sensitivity, are expected to detect 0.0021\,\% and 0.01\,\% of the FRB population respectively.  These results are summarized in Fig.\ \ref{fig:surveys} and discussed further in Section \ref{sec:galactic_scattering}.

\begin{deluxetable}{crrr|rr|rrrr}
\tablecaption{Sensitivity of the DSA-110, CRAFTS, and CHORD surveys to scattering geometries that can localize scattering to within the FRB host galaxy.  $f_\mathrm{obs}$ and BW are, respectively, the center frequency and observing bandwidth of the survey. \label{tbl:surveys}
}
\tablehead{\colhead{Survey}  & \colhead{$f_\mathrm{obs}$} & \colhead{BW} & \colhead{$\tau_\mathrm{fid}$} & \multicolumn{2}{c}{all redshifts} & \multicolumn{4}{c}{$z<0.2$}\\
\colhead{} & \colhead{} & \colhead{} & \colhead{} & \colhead{$\theta_\mathrm{res,MW}$ }  & \colhead{$f_\mathrm{sky}$} & \colhead{$\theta_\mathrm{res,MW}$ }  & \colhead{$f_\mathrm{sky}$} & \colhead{$f_z$} & \colhead{$f_\mathrm{sky} f_z$}}
\startdata
DSA-110     & 1405\,MHz &  250\,MHz &  1.0\,ms &  44\,nas &  1.2\,\% &  133\,nas & 17.4\,\% & 24\,\% &  4.2\,\%\\
DSA-2000    &  900\,MHz &  400\,MHz &  7.1\,ms & 117\,nas & 18.1\,\% &  354\,nas & 67.5\,\% &  8\,\% &  5  \,\%\\
DSA-2000    & 1800\,MHz &  400\,MHz &  0.3\,ms &  25\,nas &  0.3\,\% &   77\,nas &  7.7\,\% &  8\,\% &  0.6\,\%\\
DSA-2000    & 1350\,MHz & 1300\,MHz &  1.2\,ms &  48\,nas &  2.4\,\% &  145\,nas & 19.4\,\% &  8\,\% &  2  \,\%\\
ASKAP CRAFT &  870\,MHz &  336\,MHz &  8.1\,ms & 126\,nas & 21.0\,\% &  382\,nas & 67.2\,\% & 28\,\% & 19  \,\%\\
ASKAP CRAFT & 1320\,MHz &  336\,MHz &  1.3\,ms &  50\,nas &  4.5\,\% &  152\,nas & 21.4\,\% & 28\,\% &  5.9\,\%\\
CHORD       &  500\,MHz &  400\,MHz & 94.2\,ms & 425\,nas & 97.7\,\% & 1291\,nas & 99.3\,\% & 15\,\% & 15  \,\%\\
CHORD       & 1300\,MHz &  400\,MHz &  1.4\,ms &  52\,nas &  3.2\,\% &  158\,nas & 22.2\,\% & 15\,\% &  3.3\,\%\\
CHORD       &  900\,MHz & 1200\,MHz &  7.1\,ms & 117\,nas & 18.9\,\% &  355\,nas & 67.2\,\% & 15\,\% &  9.9\,\% 
\enddata
\end{deluxetable}

\end{document}